\documentclass{aastex631}
\usepackage{subfigure}

\shorttitle{AASTeX v6.3.1 Sample article}
\shortauthors{Wang et al.}
\graphicspath{{./}{figures/}}

\begin{document}
\title{The molecular clouds in a section of the third Galactic quadrant: observational properties and chemical
abundance ratio between CO and its isotopologues.}

\author[0000-0001-8923-7757]{Chen Wang}
\affiliation{National Astronomical Observatories, Chinese Academy of Sciences, 20A Datun Road, Chaoyang District, Beijing 100101, People's Republic of China}
\affiliation{Purple Mountain Observatory, Chinese Academy of Sciences, 10 Yuanhua Road, Qixia District, Nanjing 210033, People's Republic of China}
\affiliation{Key Laboratory of Radio Astronomy and Technology, Chinese Academy of Sciences, A20 Datun Road, Chaoyang District, Beijing 100101, People's Republic of China}

\author[0000-0003-1714-0600]{Haoran Feng}
\affiliation{Purple Mountain Observatory, Chinese Academy of Sciences, 10 Yuanhua Road, Qixia District, Nanjing 210033, People's Republic of China}
\affiliation{Key Laboratory of Radio Astronomy and Technology, Chinese Academy of Sciences, A20 Datun Road, Chaoyang District, Beijing 100101, People's Republic of China}
\affiliation{University of Science and Technology of China, Chinese Academy of Sciences, Hefei 230026, People's Republic of China}

\author[0000-0001-7768-7320]{Ji Yang}
\affiliation{Purple Mountain Observatory, Chinese Academy of Sciences, 10 Yuanhua Road, Qixia District, Nanjing 210033, People's Republic of China}
\affiliation{Key Laboratory of Radio Astronomy and Technology, Chinese Academy of Sciences, A20 Datun Road, Chaoyang District, Beijing 100101, People's Republic of China}

\author[0000-0003-3151-8964]{Xuepeng Chen}
\affiliation{Purple Mountain Observatory, Chinese Academy of Sciences, 10 Yuanhua Road, Qixia District, Nanjing 210033, People's Republic of China}
\affiliation{Key Laboratory of Radio Astronomy and Technology, Chinese Academy of Sciences, A20 Datun Road, Chaoyang District, Beijing 100101, People's Republic of China}

\author[0000-0002-0197-470X]{Yang Su}
\affiliation{Purple Mountain Observatory, Chinese Academy of Sciences, 10 Yuanhua Road, Qixia District, Nanjing 210033, People's Republic of China}
\affiliation{Key Laboratory of Radio Astronomy and Technology, Chinese Academy of Sciences, A20 Datun Road, Chaoyang District, Beijing 100101, People's Republic of China}

\author[0000-0003-4586-7751]{Qing-Zeng Yan}
\affiliation{Purple Mountain Observatory, Chinese Academy of Sciences, 10 Yuanhua Road, Qixia District, Nanjing 210033, People's Republic of China}
\affiliation{Key Laboratory of Radio Astronomy and Technology, Chinese Academy of Sciences, A20 Datun Road, Chaoyang District, Beijing 100101, People's Republic of China}

\author[0000-0002-7489-0179]{Fujun Du}
\affiliation{Purple Mountain Observatory, Chinese Academy of Sciences, 10 Yuanhua Road, Qixia District, Nanjing 210033, People's Republic of China}
\affiliation{University of Science and Technology of China, Chinese Academy of Sciences, Hefei 230026, People's Republic of China}
\affiliation{Key Laboratory of Radio Astronomy and Technology, Chinese Academy of Sciences, A20 Datun Road, Chaoyang District, Beijing 100101, People's Republic of China}

\author[0000-0002-8051-5228]{Yuehui Ma}
\affiliation{Purple Mountain Observatory, Chinese Academy of Sciences, 10 Yuanhua Road, Qixia District, Nanjing 210033, People's Republic of China}
\affiliation{Key Laboratory of Radio Astronomy and Technology, Chinese Academy of Sciences, A20 Datun Road, Chaoyang District, Beijing 100101, People's Republic of China}

\author[0000-0002-9649-8549]{Jiajun Cai}
\affiliation{College of Science \& Center for Astronomy and Space Sciences, China Three Gorges University, Yichang 443002, People's Republic of China}

\correspondingauthor{Ji Yang}
\email{jiyang@pmo.ac.cn}



\begin{abstract}
We compare the observational properties between {$^{12}$CO}, {$^{13}$CO}, and {C$^{18}$O} and summarize the observational parameters based on 7069 clouds sample from the Milky Way Imaging Scroll Painting (MWISP) CO survey in a section of the third Galactic quadrant. We find that the {$^{13}$CO} angular area ($A_{\rm ^{13}CO}$) generally increases with that of {$^{12}$CO} ($A_{\rm  ^{12}CO}$), and the ratio of $A_{\rm ^{13}CO}$ to $A_{\rm  ^{12}CO}$ is 0.38 by linear fitting. We find that the {$^{12}$CO} and {$^{13}$CO} flux are tightly correlated as $F_{\rm ^{13}CO}~=~0.17~ F_{\rm ^{12}CO}$ with both fluxes calculated within the {$^{13}$CO}-bright region. This indicates that the abundance $X_{\rm ^{13}CO}$ is a constant to be 6.5$^{+0.1}_{-0.5}$ $\times 10^{-7}$ for all samples under assumption of local thermodynamic equilibrium (LTE). Additionally, we observed that the X-factor is approximately constant in large sample molecular clouds. Similarly, we find $F_{\rm C^{18}O}~=~0.11~F_{\rm ^{13}CO}$ with both fluxes calculated within {C$^{18}$O}-bright region, which indicates that the abundance ratios ${X_{\rm ^{13}CO}/X_{\rm C^{18}O}}$ stays the same value 9.7$^{+0.6}_{-0.8}$ across the molecular clouds under LTE assumption. The linear relationships of $F_{\rm ^{12}CO}$ vs. $F_{\rm ^{13}CO}$ and $F_{\rm ^{13}CO}$ vs. $F_{\rm C^{18}O}$ hold not only for the {$^{13}$CO}-bright region or {C$^{18}$O}-bright region, but also for the entire molecular cloud scale with lower flux ratio. The abundance ratio ${X_{\rm ^{13}CO}/X_{\rm C^{18}O}}$ inside clouds shows a strong correlation with column density and temperature. This indicates that the ${X_{\rm ^{13}CO}/X_{\rm C^{18}O}}$ is dominated by a combination of chemical fractionation, selectively dissociation, and self-shielding effect inside clouds.

\end{abstract}

\keywords{ISM: molecules - ISM: clouds - ISM: abundances}
\section{Introduction}
\label{Sec:1}
Molecular clouds (MCs; \citealt{2001ApJ...547..792D}) are an important component of the interstellar medium and serve as the birthplace of stars.
CO and its isotopologues, {$^{13}$CO} and {C$^{18}$O} (J = 1-0), are effective tracers of MCs, providing valuable physical and chemical information under the assumption of local thermodynamic equilibrium (LTE; \citealt{2008ApJ...679..481P}, \citealt{2009tra..book.....W}, \citealt{2010ApJ...721..686P}). In general, the column density of molecular hydrogen can be estimated from their abundance (\citealt{2000Obs...120..289R,1998AJ....116..336N}), with $N$(H$_{2}$)/$N$({$^{13}$CO}) = 7 $\times$ $10^{5}$ and $N$(H$_{2}$)/$N$({C$^{18}$O}) = 7 $\times$ $10^{6}$. However, is the abundance of these tracers the same in all molecular clouds? Or does their abundance systematically change during the evolution of molecular clouds?

These questions have garnered significant attention in recent decades, and observational and theoretical efforts have been made to address them. \citet{1988ApJ...334..771V} demonstrated that the isotopic chemistry of CO is influenced by different physical processes and chemical reactions in various regions. For instance, CO rare isotopologues associated with OB stars or H{\scriptsize~II} regions are selectively dissociated by FUV (far-ultraviolet) emission more effectively than CO due to differences in self-shielding effects (\citealt{1988ApJ...334..771V}, \citealt{1996A&A...308..535W}, \citealt{2007A&A...476..291L}, \citealt{2009A&A...503..323V}, \citealt{2014A&A...564A..68S,2015ApJS..217....7S}).

On the other hand, the abundance ratio of CO isotopologues in dark clouds (DCs) or cold cores is close to the terrestrial value (\citealt{1992A&ARv...4....1W}), as these clouds lack FUV emission and have lower temperatures. This suggests that there are at least two types of molecular clouds with different CO isotopic abundance ratios. In fact, case studies of MCs and simulation studies have confirmed this. The impact of selective photodissociation and chemical fractionation of CO on the ${X_{\rm ^{12}CO}/X_{\rm ^{13}CO}}$ isotopic ratio has been examined by \citet{2014MNRAS.445.4055S}, and they found a close correlation between the ratio and both the $^{12}$CO and $^{13}$CO column densities.

In observational case studies, \citet{1992A&ARv...4....1W} show a gradient in the {$^{12}$C/$^{13}$C} and {$^{16}$O/$^{18}$O} ratios with galactocentric distance, and get the terrestrial ${X_{\rm ^{13}CO}/X_{\rm C^{18}O}}$ value is 5.5. Furthermore, \citet{2012ApJ...756...76W} found that the mean abundance ratio of ${X_{\rm ^{13}CO}/X_{\rm C^{18}O}}$ in 674 Planck cold clumps from the Early Cold Core Catalogue (ECC) is 7  with LTE assumption, based on a survey of the J = 1-0 transitions of {$^{12}$CO}, {$^{13}$CO}, and {C$^{18}$O} using the Purple Mountain Observatory (PMO) 13.7 m telescope. In the last decade, the effect of selective photodissociation have received more attention and discussion. \citet{2014A&A...564A..68S} found that the mean ${X_{\rm ^{13}CO}/X_{\rm C^{18}O}}$ ratio in nearly edge-on photon-dominated regions is 16.47, which is much larger than the terrestrial value of 5.5 (\citealt{1992A&ARv...4....1W}), while the ${X_{\rm ^{13}CO}/X_{\rm C^{18}O}}$ ratio in other regions was found to be 12.29 using the Nobeyama 45 m telescope in the J = 1-0 transitions of CO isotopologues with LTE assumption. Additionally, \citet{2015ApJ...805...58K} found significant spatial variations in the abundances of {$^{13}$CO} and {C$^{18}$O} in the southeastern part of the California MC that are correlated with variations in gas temperature surveyed in the {$^{12}$CO}, {$^{13}$CO}, and {C$^{18}$O}(J = 2-1) with standard LTE analysis. \citet{2021A&A...645A..26R} found that chemical effects play a significant role, by enhancing the {$^{13}$CO} abundance (fractionation) or destroying {C$^{18}$O} (photodissociation) when the ${X_{\rm ^{13}CO}/X_{\rm C^{18}O}}$ upper bound is close to 50 in Orion B molecular cloud by the Cramer Rao bound technique. These results indicate that ${X_{\rm ^{13}CO}/X_{\rm C^{18}O}}$ will change significantly with the presence of heating sources inside molecular clouds.  Importantly, our previous work, based on large-scale (58.5 deg$^2$) observations of {$^{12}$CO}, {$^{13}$CO}, and {C$^{18}$O} $(1-0)$ transitions with LTE assumption towards the Gemini OB1 Molecular Cloud Complex, found that clouds can be divided into two types based on ${X_{\rm ^{13}CO}/X_{\rm C^{18}O}}$ ratios (\citealt{2019ApJS..243...25W}). These two types of clouds exhibit systematic differences in their physical properties, such as excitation temperature and column density.

Although differences in isotopic abundance between molecular clouds have been observed, it is not known whether these differences are systematic due to the limited number of samples studied, and the exact physical explanation behind them remains unclear. Studying isotopic abundance in large samples can provide a way to address this problem.

The data in this study covering an area of 250 square degrees located in the third quadrant of the Milky Way. This paper is the third in a series of studies on large samples of molecular clouds in this region. In Paper I (\citealt{2023AJ....165..106W}), the data were described in detail, and 7069 clouds were identified based on the DBSCAN algorithm in the {$^{12}$CO} data. Additionally, we identified 1197 clouds with {$^{13}$CO} emission and 32 clouds with {C$^{18}$O} emission based on the stacking bump algorithm in the catalog of the 7069 clouds. In Paper II (\citealt{2022ApJS..262...16M}), column density maps were obtained for an unbiased sample of 120 molecular clouds in the data, and a relation between the dispersion of normalized column density and the sonic Mach number of molecular clouds was found based on the same data and catalog.

In this paper, we study the fundamental properties of molecular clouds following the catalog in Paper I. The observations and data reduction are described in Section~\ref{Sec:2}. In Section~\ref{Sec:3}, we present the interesting results, which include a statistical study of the observational properties of molecular clouds, optical depth, and flux results. The results of the abundance $X_{\rm ^{13}CO}$ and abundance ratio $X_{\rm ^{13}CO}/X_{\rm C^{18}O}$ are presented in Section~\ref{Sec:4} and Section~\ref{Sec:5}, respectively. We will discuss some important issues in Section~\ref{Sec:6}, and the conclusion is given in Section~\ref{Sec:7}.

\section{Data and catalog}
\label{Sec:2}

The data of the J = 1-0 transitions of $^{12}$CO, $^{13}$CO, and C$^{18}$O were obtained with the Purple Mountain Observatory (PMO) 13.7 m millimeter telescope (\citealt{4534591}, \citealt{2012ITTST...2..593S}). The data cover a section in the third Galactic quadrant ($195^{\circ} < l < 220^{\circ}$, $|b| < 5^{\circ}$) and are part of the Milky Way Imaging Scroll Painting (MWISP) project, which is a multi-line Galactic plane survey in CO and its isotopic transitions. The mean noise $\sigma$ is $\sim$0.45~K for {$^{12}$CO} with 0.16 km s$^{-1}$ velocity resolution, and $\sim$0.25~K for {$^{13}$CO} and {C$^{18}$O} with 0.17 km s$^{-1}$ velocity resolution (\citealt{2023AJ....165..106W}).

Paper I has described the data in detail and has shown that 7069 clouds were identified based on the DBSCAN algorithm in the $^{12}$CO data (\citealt{e1abb5edb3604fb39e22265c59b8815c}). We have also identified 1197 clouds with $^{13}$CO emission and 32 clouds with C$^{18}$O emission based on the stacking bump algorithm in the 7069 cloud catalog, respectively. In this paper, we study the fundamental properties of molecular clouds following the catalog in Paper I.

\section{Observational properties between {$^{12}$CO}, {$^{13}$CO} and {C$^{18}$O}}
\label{Sec:3}

  \subsection{peak intensity}
  \label{Sec:31}
Observationally, the specific form of the radiative equation in the low-temperature approximation (\citealt{1997ApJ...476..781B}) is given by
\begin{equation}
T_{\rm mb} = \eta[J(T_{\rm ex})-J(T_{\rm bg})](1 - \exp (-\tau)),
\end{equation}
where $T_{\rm mb}$ is the observed brightness temperature, $\eta$ is the beam filling factor (assumed to be 1, as discussed in Section~\ref{Sec:62}), and $J(T)$ is a function of temperature $T$, given by $\rm J(T)= h\nu/k \times (exp(h\nu/kT)-1)^{-1}$. $T_{\rm ex}$ is the excitation temperature, $T_{\rm bg}$ is the background temperature, and $\tau$ is the optical depth.

Assuming the molecular clouds (MCs) are in LTE condition, we can obtain the excitation temperature using the $^{12}$CO (J=1$-$0) emission line, which is generally known to be optically thick (e.g., Section~\ref{Sec:32}). The calculation is based on the following formula (\citealt{2000Obs...120..289R,1998AJ....116..336N}):

\begin{equation}
T_{\rm ex} = \frac{5.53}{{\rm ln} (1+\frac{5.53 K}{T_{\rm mb} (^{12}{\rm CO})+0.819 K})} K,
\end{equation}

In order to avoid the influence of noise, we selected pixels based on the signal-to-noise ratio (S/N) for our statistical research. All of the pixels are from the cloud catalog described in Paper I. We considered only pixels with a spectral maximum greater than 4~$\sigma$ ($\sim$ 1.8 K for $^{12}$CO and $\sim$ 1 K for $^{13}$CO and C$^{18}$O). Additionally, we considered only velocity-integrated intensity pixels with an intensity greater than five times the intensity noise threshold ($\sigma_i$ = $\sigma\sqrt{ N_{\rm channels}\Delta v_i}$, where $N_{\rm channels}$ is the number of channels in the spectrum).

\begin{figure}
   \centering
   \includegraphics[width=\textwidth]{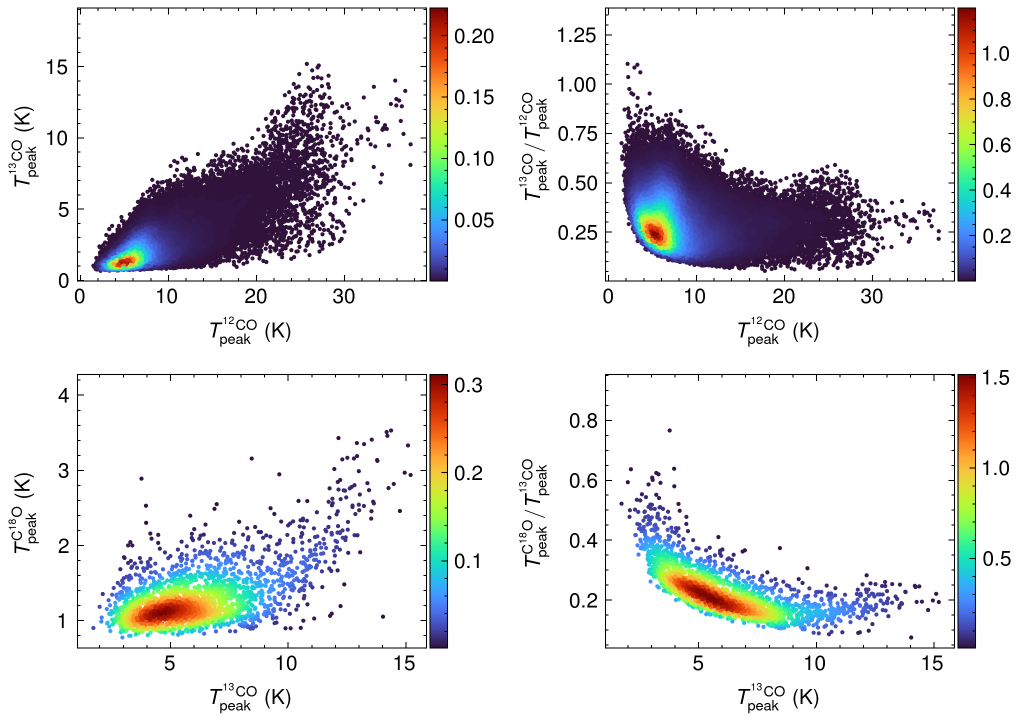}
   \caption{Peak intensity distribution and peak intensity ratio pixel by pixel measured in {$^{12}$CO}, {$^{13}$CO}, and{C$^{18}$O} catalog. The color coding represents the distribution of the probability density functions of the pixels with $^{13}$CO emission (2D-PDFs), which are calculated by utilizing kernel density estimation through Gaussian kernels in the Python package scipy.stats.Gaussian\_kde(\url{https://docs.scipy.org/doc/scipy/reference/generated/scipy.stats.gaussian_kde.html})}
    \label{Fig:Tpeak}
\end{figure}

Figure~\ref{Fig:Tpeak} shows $T_{\rm peak}^{\rm ^{13}CO}$ vs. $T_{\rm peak}^{\rm ^{12}CO}$ measured pixel by pixel in 1197 clouds that have $^{13}$CO emission. The maximum $T_{\rm peak}^{\rm ^{12}CO}$ and $T_{\rm peak}^{\rm ^{13}CO}$ are 37.5 K and 15.2 K, respectively (Table~\ref{table1}). 88.3\% of the pixels have $T_{\rm peak}^{\rm ^{13}CO}$ and $T_{\rm peak}^{\rm ^{12}CO}$ in the range of [1, 5] K and [1, 10] K, respectively.

Figure~\ref{Fig:Tpeak} presents ${T_{\rm peak}^{\rm ^{13}CO}/T_{\rm peak}^{\rm ^{12}CO}}$ against $T_{\rm peak}^{\rm ^{12}CO}$ measured pixel by pixel in 1197 clouds that have $^{13}$CO emission. The dispersion of the peak ratio ${T_{\rm peak}^{\rm ^{13}CO}/T_{\rm peak}^{\rm ^{12}CO}}$ decreases with $T_{\rm peak}^{\rm ^{12}CO}$, and there is a bump distribuion in the range of [20, 30] K.

The relationship between $T_{\rm peak}^{\rm ^{13}CO}$ and $T_\mathrm{\rm peak}^{\rm C^{18}O}$ in a catalog of 32 clouds that have C$^{18}$O emission is shown in Figure~\ref{Fig:Tpeak}. The maximum $T_{\rm peak}^{\rm C^{18}O}$ is 3.53 K. 81\% of the pixels have $T_{\rm peak}^{\rm C^{18}O}$ in the range of [0.7, 1.5] K. This shows that with limited sensitivity truncation, the S/N ratios of {$^{12}$CO}, {$^{13}$CO}, and C$^{18}$O obviously decrease sequentially. In the statistical analysis of C$^{18}$O, the selection effect caused by detection sensitivity should be carefully considered. Similar to the distribution of ${T_{\rm peak}^{^{13}\rm CO}/T_{\rm peak}^{^{12}\rm CO}}$, Figure~\ref{Fig:Tpeak} shows that the ratio ${T_{\rm peak}^{C^{18}\rm O}/T_{\rm peak}^{^{13}\rm CO}}$ decreases with $T_{\rm peak}^{\rm ^{13}CO}$ and stays in the range of [0.1, 0.3] in the high value region of $T_{\rm peak}^{\rm ^{13}CO}$.

  \subsection{optical depth}
  \label{Sec:32}
On the other hand, the optical depth of $^{13}$CO and C$^{18}$O emission can be derived as follows (\citealt{2010ApJ...721..686P})

\begin{equation}
\tau (^{13}{\rm CO})=-\ln \left[1-\frac{T_{\rm mb} (^{13}{\rm CO})}{5.29} \left(\left[e^{5.29/T_{\rm ex}}-1\right]^{-1}-0.164\right)^{-1}\right]
\end{equation}
\begin{equation}
\tau ({\rm C^{18}O})=-\ln \left[1-\frac{T_{\rm mb}({\rm C^{18}O})}{5.27} \left(\left[e^{5.27/T_{\rm ex}}-1\right]^{-1}-0.166\right)^{-1}\right]
\end{equation}

\begin{figure}
   \centering
   \subfigure[]{
   \label{Fig:figure-A}
   \includegraphics[width=0.3\textwidth]{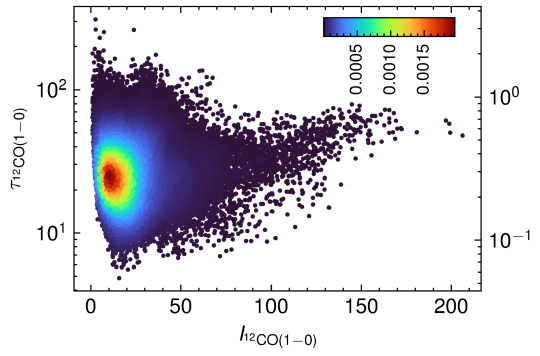}}
   \subfigure[]{
   \label{Fig:figure-B}
   \includegraphics[width=0.3\textwidth]{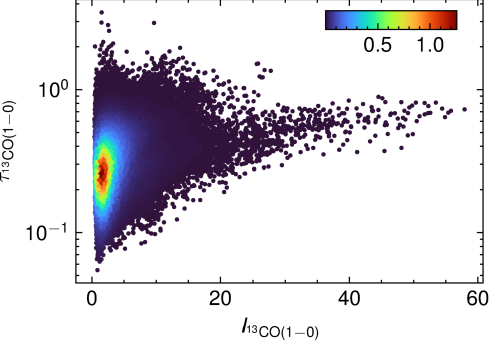}}
   \subfigure[]{
   \label{Fig:figure-C}
   \includegraphics[width=0.3\textwidth]{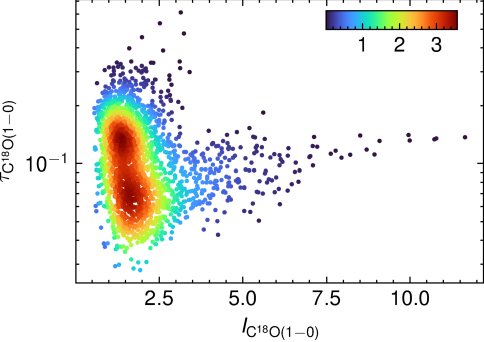}}

   \caption{Optical depth $\tau $ vs. velocity-integrated intensity I pixel by pixel measured in {$^{12}$CO}, {$^{13}$CO}, and{C$^{18}$O} catalog. The color coding are same with Figure~\ref{Fig:Tpeak}.}
    \label{Fig:tau}
\end{figure}

The variation of $\tau (^{13}{\rm CO})$ with $^{13}$CO integrated intensity is shown in Figure~\ref{Fig:tau}. The dispersion of $\tau (^{13}{\rm CO})$ decreases rapidly with $^{13}$CO integrated intensity. For about 70\% of pixels, $\tau (^{13}{\rm CO})$ falls within the range of [0.1, 0.4].

Assuming typical relative abundances for the CO isotopologues in the local ISM ($\rm{X_{^{12}CO}/X_{^{13}CO}}$=89, \citealt{1980afcp.book.....L}), the opacity $\tau (^{12}{\rm CO})$ can be easily obtained from $\tau (^{13}{\rm CO})$ by multiplying the ratio $\rm{X_{^{12}CO}/X_{^{13}CO}}$ (Figure~\ref{Fig:tau}). It shows that almost all $^{12}$CO is optically thick ($\tau (^{12}{\rm CO})$ $>$ 1), with 77\% of pixels falling within the $\tau (^{12}{\rm CO})$ range of [10,40]. This indicates that the approximate relation of equation (2) is valid in most clouds. It is important to note that $\tau (^{12}{\rm CO})$ is derived from $\tau (^{13}{\rm CO})$, so we cannot obtain $\tau (^{12}{\rm CO})$ values in some diffuse regions where $^{13}{\rm CO}$ emission is not detected, and these $\tau (^{12}{\rm CO})$ results are not included in our diagram (Figure~\ref{Fig:tau}).

According to equation (4), the $\tau ({\rm C^{18}O})$ results for 32 clouds catalog with C$^{18}$O emission are shown in Figure~\ref{Fig:tau}. The dispersion of $\tau ({\rm C^{18}O})$ decreases with C$^{18}$O emission and remains within the range of [0.05, 0.1]. The value of $\tau ({\rm C^{18}O})$ is concentrated between 0.04 and 0.2 for more than 90\% of pixels, which is much smaller than the values of $\tau (^{12}{\rm CO})$ and $\tau (^{13}{\rm CO})$. This suggests that in most clouds, C$^{18}$O is optically thin.

  \subsection{integrated intensity}
  \label{Sec:33}

\begin{figure}
   \centering
   \includegraphics[width=\textwidth]{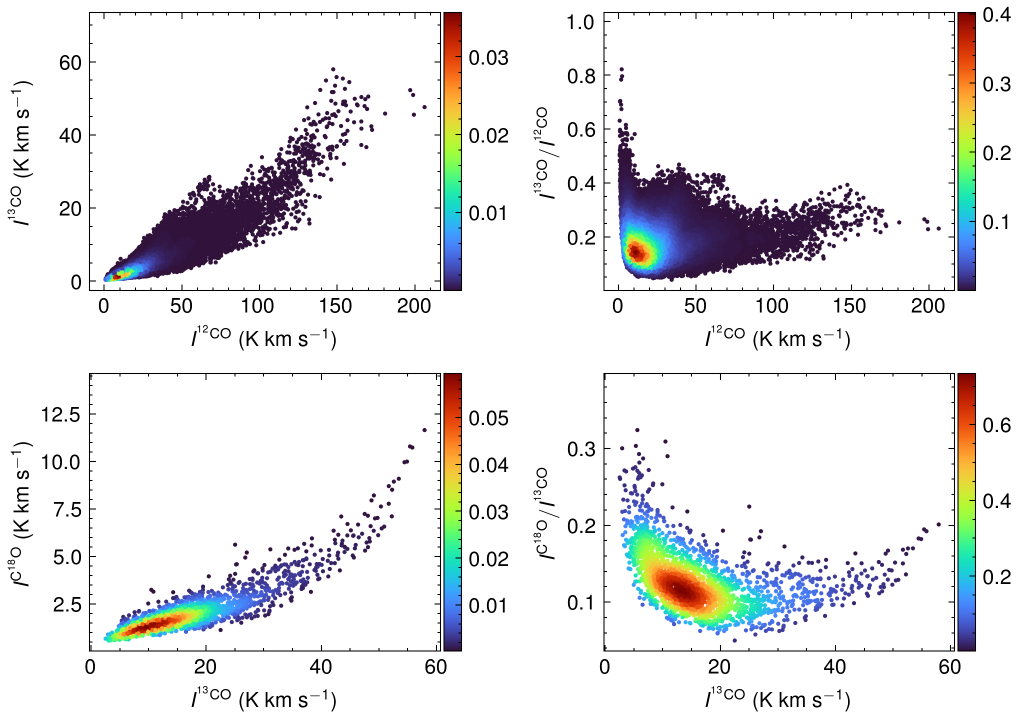}
   \caption{Intensity distribution between {$^{12}$CO}, {$^{13}$CO}, and{C$^{18}$O} pixel by pixel. The color coding are same with Figure~\ref{Fig:Tpeak}.}
    \label{Fig:intensity}
\end{figure}

Figure~\ref{Fig:intensity} shows relationship between integrated intensity of {$^{12}$CO} and {$^{13}$CO} in 1197 clouds catalog. The {$^{12}$CO} intensity in the range of [0.86, 206.25] K km s$^{-1}$, and {$^{13}$CO} in the range of [0.38, 57.93] K km s$^{-1}$ (Table~\ref{table1}). 80.8 $\%$ pixels have intensity of {$^{12}$CO} and {$^{13}$CO} in range of [1,~30]~K km s$^{-1}$ and [0.4,~5]~K km s$^{-1}$, respectively. {$^{13}$CO} intensity is increasing with {$^{12}$CO} emission, the slope (${I^{\rm ^{13}CO}/I^{^{12}\rm CO}}$) slowly increases from 0.15 to 0.25 with the {$^{12}$CO} intensity. The change in the growth slope may be caused by  two reasons, one is the increase in the optical depth of {$^{12}$CO} in the high intensity region which result in the decrease of {$^{12}$CO} intensity, and the other is the increase in {$^{13}$CO} intensity due to the self-shielding effect.

Figure~\ref{Fig:intensity} presents the ratio ${I^{\rm ^{13}CO}/I^{^{12}\rm CO}}$ in 1197 clouds catalog. The scatter of the ratio is decreasing with $\rm I^{^{12}\rm CO}$ and concentrated on $ \sim $0.3 in the high intensity region. Different with high intensity region ditribution, 73.38 $\%$ pixels have ratio ${I^{\rm ^{13}CO}/I^{^{12}\rm CO}}$ in range of [0, 0.2].

Relationship between integrated intensity of {$^{13}$CO} and {C$^{18}$O} in 32 clouds catalog is presented in Figure~\ref{Fig:intensity}. The {C$^{18}$O} intensity in the range of [0.57, 11.65] K km s$^{-1}$. Similar with {$^{12}$CO} against {$^{13}$CO}, {C$^{18}$O} intensity is increasing with {$^{13}$CO} emission, and the slope becomes steeper after {$^{13}$CO} intensity greater than 40 K km s$^{-1}$. The ratio ${I^{\rm C^{18}O}/I^{\rm ^{13}CO}}$ is decreasing with $I^{^{13}\rm CO}$ in low intensity region but increasing with $I^{^{13}\rm CO}$ in high intensity region (Figure~\ref{Fig:intensity}). The ratio ${I^{\rm C^{18}O}/I^{\rm ^{13}CO}}$ is concentrated on [0.05, 0.15]. As discussed in Sect~\ref{Sec:4}, the change in slope may be related to selective far-UV photodissociation (\citealt{1997ARA&A..35..179H}) and self-shielding effects (\citealt{1988ApJ...334..771V}, \citealt{1996A&A...308..535W}, \citealt{2007A&A...476..291L}, \citealt{2009A&A...503..323V}, \citealt{2014A&A...564A..68S,2015ApJS..217....7S}).

  \subsection{angular area}
  \label{Sec:34}
\begin{figure}
   \centering
   \includegraphics[width=\textwidth]{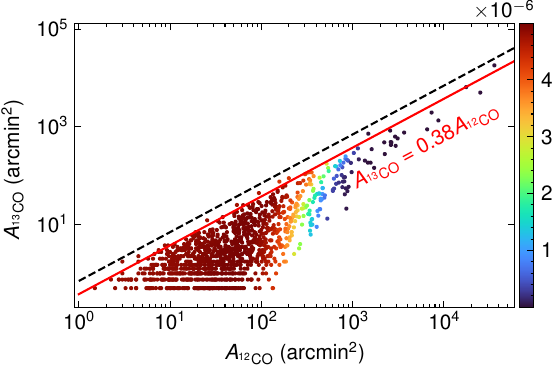}
   \caption{{$^{12}$CO} area vs. {$^{13}$CO} area cloud by cloud measured in 1197 clouds catalog. The color coding are same with Figure~\ref{Fig:Tpeak}. The red line represents a linear fit with a ratio of 0.38. The fitting calculated by utilizing the Ordinary least square (OLS) function in the Python package Statsmodels (\url{https://www.statsmodels.org/dev/generated/statsmodels.regression.linear_model.OLS.html}), assuming a zero intercept and no weighting. The dashed black line is the limit with a ratio of 0.7 proposed by \citet{2022ApJS..261...37Y}.
    }
    \label{Fig:A13-A12}
\end{figure}

Figure~\ref{Fig:A13-A12} shows the relationship between the angular areas of {$^{12}$CO} and {$^{13}$CO} emission. The angular area of {$^{13}$CO} ($A_{^{13}\rm CO}$) increases with $A_{^{12}\rm CO}$, and its dispersion decreases. Interestingly, there appears to be a upper ratio of 0.7, with the area ratio of {$^{13}$CO} to {$^{12}$CO} approaching 0.7 as the cloud area increases. \citet{2022ApJS..261...37Y} also found that the area of {$^{13}$CO} emission in a molecular cloud generally does not exceed 70\% of the {$^{12}$CO} emission area based on the second Galactic quadrant from the same MWISP project. Figure~\ref{Fig:A13-A12} confirm that our sample is also within this limit of 0.7 ratio. Through linear fitting of the data, the angular area ratio of {$^{12}$CO} and {$^{13}$CO} can be obtained as 0.38. The fitting calculated by utilizing the Ordinary least square (OLS) function in the Python package Statsmodels, assuming a zero intercept and no weighting. It should be noted that our fit is performed in a linear space, whereas the Figure~\ref{Fig:A13-A12} shows a distribution in a logarithmic space. Therefore, in Figure~\ref{Fig:A13-A12}, there are some molecular clouds at the bottom left that deviate from the fitted line with a large difference, but in the actual linear space, they differ little from the fitted line. The value of ratio 0.38 may be related to the sensitivity of the data, which is worth discussing in detail in our next work (Wang et al. in prep). \citet{2020ApJS..246....7S} found that ratios of $A_{^{13}\rm CO}$ to $A_{^{12}\rm CO}$ varies from 0.04 to 0.26 in different Galactic
arms based on the same data of the MWISP over l = [$129^{\circ}.75$, $140^{\circ}.25$]. This suggests that the distance factor has a significant effect on the ratio. It is noted that the areas of {$^{12}$CO} and {$^{13}$CO} emission have a large dispersion in the region where $A_{^{13}\rm CO}$ is less than 10$^3$ arcmin$^2$. This could be caused by filling factor or sensitivity, as discussed in Sect~\ref{Sec:62}. The beam filling factor is mainly correlated with the angular size,which makes the {$^{13}$CO} signal more difficult to detect in smaller molecular clouds. Therefore, for small angular area molecular clouds, the area detection of {$^{13}$CO} is seriously underestimated.This also explains why the $A_{^{13}\rm CO}$ of molecular clouds with smaller angular area is basically much smaller than the result of linear fitting.

  \subsection{flux}
  \label{Sec:35}
In order to clarify the molecular cloud regions discussed, we divide the molecular cloud into three parts based on the CO isotopologues emission characteristics: only {$^{12}$CO}-ongly, {$^{13}$CO}-bright, and {C$^{18}$O}-bright. The {$^{12}$CO}-ongly represents a region with only {$^{12}$CO} emission and no {$^{13}$CO} and {C$^{18}$O} detected. The {$^{13}$CO}-bright represents a region with {$^{12}$CO} and {$^{13}$CO}  emission but on {C$^{18}$O} detected. Finally, the {C$^{18}$O}-bright is the region which all of CO isotopologues detected.

\begin{figure}
   \centering

   \includegraphics[width=\textwidth, angle=0,clip=true,keepaspectratio=true,trim=0 0 0 0mm]{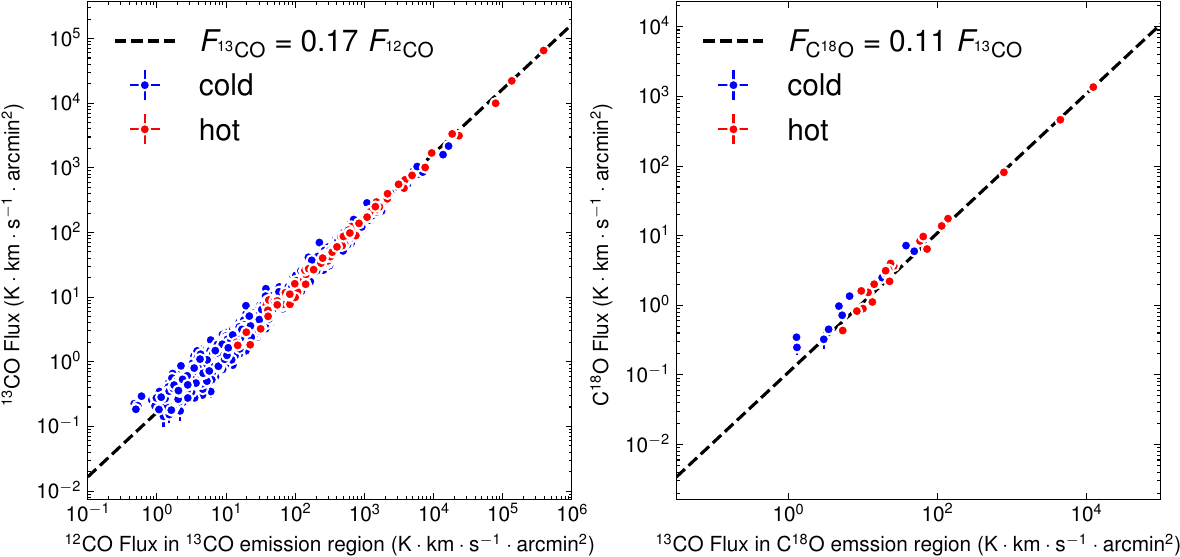}
   \caption{Left:$\rm {Flux(^{12} CO)}$ vs $\rm {Flux(^{13} CO)}$ measured cloud by cloud in in 1197 clouds catalog. $\rm {Flux(^{12} CO)}$ and $\rm {Flux(^{13} CO)}$ are calculated toward the same areas where both {$^{12}$CO} and {$^{13}$CO} emissions are detected. Right: $\rm {Flux(^{13} CO)}$ vs $\rm {Flux( C^{18}O)}$ measured in 32 clouds catalog. $\rm {Flux(^{13} CO)}$ and $\rm {Flux( C^{18}O)}$ are calculated toward the same areas where both  {$^{13}$CO} and {C$^{18}$O} emissions are detected. The colors indicate different types of molecular clouds based on excitation temperature.
   }
    \label{Fig:flux}
\end{figure}

A strong correlation between the flux of {$^{12}$CO} and {$^{13}$CO} in {$^{13}$CO}-bright region can be found in Figure~\ref{Fig:flux}, spanning six orders of magnitude. It should be noted that the {$^{12}$CO} flux in Figure~\ref{Fig:flux} only includes the {$^{13}$CO}-bright region, but does not include the diffuse region with only {$^{12}$CO} emission. Despite the large optical depth ratios of {$^{12}$CO} to {$^{13}$CO}, it is interesting that the observed flux ratios are tightly correlated over such a wide range. We divided the molecular cloud into hot and cold clouds based on the maximum excitation temperature of 14 K. Figure~\ref{Fig:flux} shows that hot clouds have higher flux values of {$^{12}$CO} and {$^{13}$CO} compared to cold clouds, but the flux ratio remains the same.

In general, for an optically thin molecular line, the flux of the molecular line is proportional to the mass of the molecule, and the flux of {$^{12}$CO} is proportional to the total mass of the cloud, thanks to the X-factor (\citealt{2001ApJ...547..792D}, \citealt{2013ARA&A..51..207B}). The strong correlation in Figure~\ref{Fig:flux} indicates that the molecular cloud in the section of the third Galactic quadrant has a stable and uniform {$^{13}$CO} abundance value (Sect~\ref{Sec:4}), and it also suggests that the value of the X-factor is statistically stable at the overall scale of the molecular cloud (Sect~\ref{Sec:61}). Through linear fitting of the data, we found that the data perfectly fit the relation $F_{\rm ^{13}CO} = 0.17~F_{\rm ^{12}CO}$. Since the majority of pixels have a ratio ${I^{\rm ^{13}CO}/I^{^{12}\rm CO}}$ in the range of [0, 0.2] and {$^{12}$CO} intensity in the range of [0, 30] K km s$^{-1}$ (Figure\ref{Fig:intensity}), the relation $F_{\rm ^{13}CO} = 0.17~F_{\rm ^{12}CO}$ indicates that the abundance $X_{\rm ^{13}CO}$ is a constant for the entire samples (Section~\ref{Sec:4}).

Another similar strong correlation between the flux of {$^{13}$CO} and {C$^{18}$O} is also found in Figure~\ref{Fig:flux}. It should be noted that the {$^{13}$CO} flux in here only includes the region in the molecular cloud with {C$^{18}$O} emission. Similar to Figure~\ref{Fig:flux}, hot clouds have higher flux values than cold clouds, but the flux ratio remains the same. Through linear fitting of the data, we found that the data perfectly fit the relation $F_{\rm C^{18}O} = 0.11~F_{\rm ^{13}CO}$. This is consistent with Figure~\ref{Fig:intensity}, which shows that the ratio ${I^{\rm C^{18}O}/I^{\rm ^{13}CO}}$ is concentrated in the range of [0.05, 0.15]. Such consistent flux ratios suggest that the abundance ratios $X_{\rm ^{13}CO}/X_{\rm C^{18}O}$ is a constant at the molecular cloud scale.

In general, the column density of {$^{13}$CO} and {C$^{18}$O} in a molecular cloud can be calculated from the integral intensity of an optically thin molecular cloud line, if the excitation temperature of the cloud is known (\citealt{1991ApJ...374..540G,1997ApJ...476..781B}). The column density equations for {$^{13}$CO} and {C$^{18}$O} are given by:

\begin{equation}
N(^{13}{\rm CO}) = 2.42 \times 10^{14}\frac{\tau (^{13}{\rm CO})}{1-e^{-\tau (^{13}{\rm CO})}} \frac{1+0.88/T_{\rm ex}}{1-e^{-5.29/T_{\rm ex}}} \int T_{\rm mb}(^{13}{\rm CO}) dv
\end{equation}

\begin{equation}
N({\rm C^{18}O}) = 2.54 \times 10^{14}\frac{\tau ({\rm C^{18}O})}{1-e^{-\tau ({\rm C^{18}O})}} \frac{1+0.88/T_{\rm ex}}{1-e^{-5.27/T_{\rm ex}}} \int T_{\rm mb}({\rm C^{18}O}) dv
\end{equation}

So, the flux ratio $\rm F_{C^{18}O}/ F_{^{13}CO}$ can be derived as:

\begin{equation}
\frac{F_{\rm C^{18}O}}{F_{\rm ^{13}CO}} = \frac{2.42 \times 10^{14}}{2.54 \times 10^{14}} \times \frac{M(\rm C^{18}{O})}{M(^{13}{\rm CO})} \frac{\tau (^{13}{\rm CO})}{1-e^{-\tau (^{13}{\rm CO})}} \frac{1-e^{-\tau ({\rm C^{18}O})}}{\tau ({\rm C^{18}O})}
\end{equation}

Here, $M$($^{13}$CO) and $M$(C$^{18}$O) represent the masses of $^{13}$CO and C$^{18}$O in the cloud, respectively. It is observed that in the third Galactic quadrant, the abundance ratio of $X_{\rm ^{13}CO}/X_{\rm C^{18}O}$ (= $M$($^{13}$CO)/$M$(C$^{18}$O)) is a constant of  9.7$^{+0.6}_{-0.8}$ (the error is derived from the change in optical depth) on the molecular cloud scale calculated from equation (7) and the relation $F_{\rm C^{18}O} = 0.11~F_{\rm ^{13}CO}$ based on the mean optical thickness of {$^{13}$CO} and {C$^{18}$O} (Table~\ref{table1}) are 0.36 and 0.12, respectively. This value is nearly twice as high as the terrestrial ratio of 5.5 (\citealt{1992A&ARv...4....1W}). This result is consistent with the findings shown in Figure~\ref{Fig:N13abund}, which indicates that $X_{\rm ^{13}CO}/X_{\rm C^{18}O}$ inside the cloud is concentrated (90.6\%) in the interval [5, 18]. The value of $X_{\rm ^{13}CO}/X_{\rm C^{18}O}$ as 9.7$^{+0.6}_{-0.8}$ falls between the Planck cold clumps of the Early Cold Core Catalogue with a value of 7 (\citealt{2012ApJ...756...76W}), and the Orion-A giant molecular cloud (except for photon-dominated regions) with a value of 12.29 (\citealt{2014A&A...564A..68S}). Since our sample contains molecular clouds that may be in different evolutionary stages, such as cold clouds and hot clouds, this stable abundance ratio indicates that the isotope abundance ratio remains constant at the overall scale of molecular clouds during the evolution of molecular clouds. It is not unique, \citet{2017MNRAS.470..401R} and \citet{2017ApJ...840L..11S} shown that known photo-chemical effects (selective photodissociation and fractionation), cannot induce global isotopologue abundances to differ from the intrinsic, IMF-determined, isotopic abundances in star-forming galaxies.

\section{Abundance $X_{\rm ^{13}CO}$}
  \label{Sec:4}

As showed in the Sect~\ref{Sec:35}, {$^{13}$CO} column density of the cloud can be obtained with equation (5). On the other hand, the column density of {H$_{2}$} in each pixel can be obtained from {$^{12}$CO} emission thanks to the X-factor (\citealt{2001ApJ...547..792D}, \citealt{2013ARA&A..51..207B}):

\begin{equation}
N({\rm H_{2}}) = X \times I_{{12}{\rm CO}}
\end{equation}

So, the flux ratio $F_{\rm ^{13}CO} / F_{\rm ^{12}CO}$  can  be derived:

\begin{equation}
\frac{F_{\rm ^{13}CO}}{F_{\rm ^{12}CO}} =\frac{M(^{13}{\rm CO})}{M({H_{2}})}\frac{X}{2.42 \times 10^{14}} \frac{1-e^{-\tau (^{13}{\rm CO})}}{\tau (^{13}{\rm CO})} \frac{1-e^{-5.29/T_{\rm ex}}}{1+0.88/T_{\rm ex}}
\end{equation}
Here, $M(^{13}\text{CO})$ is the mass of $^{13}$CO in the cloud, and $M(\text{H}_2)$ is the mass of H$_2$ in the cloud. The ratio between $M(^{13}\text{CO})$ and $M(\text{H}_2)$ represents the abundance of $^{13}$CO, denoted as $X_{\rm ^{13}CO}$, in the cloud. As shown in Section~\ref{Sec:31}, the optical depth $\tau(^{13}\text{CO})$ of most molecular gases is concentrated around 0.1, and the peak temperature $T_{\rm peak}^{\rm ^{12}CO}$ is around 10 K, corresponding to an excitation temperature of 13.4 K. Therefore, the value of $X_{\rm ^{13}CO}$ is estimated to be $6.5^{+0.1}_{-0.5} \times 10^{-7}$ (the error is derived from the change in optical depth) based on the ${F_{\rm ^{13} CO}/F_{\rm ^{12} CO}}$ value of 0.17 and the X-factor of $(2 \times 10^{20}~\text{cm}^{-2}~(\text{K km s}^{-1})^{-1}$, \citet{2013ARA&A..51..207B}).

The findings from \citet{2005ApJ...634.1126M} and \citet{2019ApJS..243...25W} suggest that there may be variations in the $^{12}\text{C}/^{13}\text{C}$ ratio and $\rm X_{^{13}\text{CO}}$ value in different regions of molecular clouds and with Galactic distance. However, our results show that, overall, the $\rm X_{^{13}\text{CO}}$ value remains stable at a fixed value without significant distance variation, and there is no obvious relationship with the local environment and evolution. This implies that although there may be local changes in molecular cloud properties due to star-forming activities and other factors, the overall properties of molecular clouds remain relatively stable. This is consistent with the discussion on $X_{\rm ^{13}CO}/X_{\rm C^{18}O}$ in the next section, suggesting that molecular clouds hold constant $X_{\rm ^{13}CO}/X_{\rm C^{18}O}$ ratio despite local variation inside. It should be noted that this conclusion is based on the premise of LTE assumption and the assumption that the X-factor is generally true and maintains a fixed value. Meanwhile, we assume that the molecular cloud has a relatively consistent $\tau (^{13}{\rm CO})$ and excitation temperature. However, as the correlated axes will induce a potentially strong bias, we remain cautious about the result that a constant abundance $X_{\rm ^{13}CO}$ over the sample. Interestingly, the linear relationship between {$^{12}$CO} and {$^{13}$CO} flux holds not only for {$^{13}$CO}-bright region, but also for the entire molecular cloud scale (Section~\ref{Sec:64}). This will also be further studied in detail in our next work (Wang et al. in prep).

\section{Abundance ratio $X_{\rm ^{13}CO}/X_{\rm C^{18}O}$}
\label{Sec:5}

\begin{figure}
   \centering
   \includegraphics[width=\textwidth]{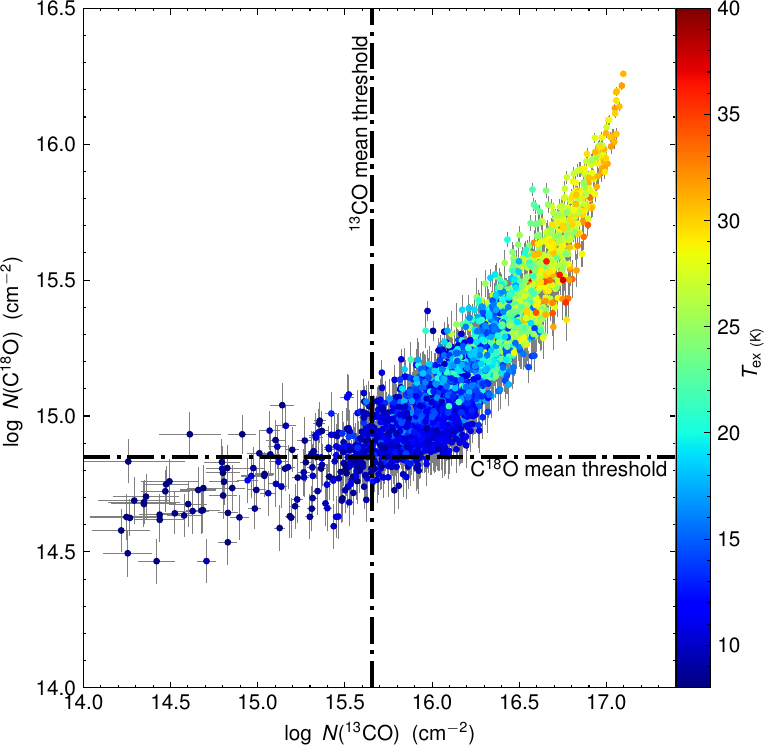}
   \caption{${N_{\rm ^{13}CO}}$ vs ${N_{\rm C^{18}O}}$ pixel by pixel measured in 32 clouds. The color coding show the excitation temperature derived from {$^{12}$CO} peak intensity. The two dashed lines show the mean threshold of integrated intensity noise, which corresponds to ${N_{\rm ^{13}CO}}$ and ${N_{\rm C^{18}O}}$ of $\rm{4.5 \times 10^{15} cm^{-2}}$  and $\rm{7.0 \times 10^{14} cm^{-2}}$, respectively. The gray lines show the error bar for each pixel.
   }
    \label{Fig:N13N18}
\end{figure}

Figure~\ref{Fig:N13N18} shows the pixel-by-pixel plot of ${N_{\rm ^{13}CO}}$ versus ${N_{\rm C^{18}O}}$ for the 32 cloud samples. The pixel selection is the same as the results presented in Sect~\ref{Sec:31}. It can be observed from Figure~\ref{Fig:N13N18} that the distributions of $N(^{13}{\rm CO})$, $N({\rm C^{18}O})$, and $T_{\rm ex}$ are closely related. $N({\rm C^{18}O})$ and $T_{\rm ex}$ increase with increasing $N(^{13}{\rm CO})$, and all three show a simple, monotonically increasing correlation. Since the intensity noise $\sigma_i$ is related to the velocity range of integration, and the velocity range of each molecular cloud is quite different, the noise $\sigma_i$ of each pixel is also different. We calculated the mean threshold of integrated intensity noise, which corresponds to ${N_{\rm ^{13}CO}}$ and ${N_{\rm C^{18}O}}$ of $\rm{4.5 \times 10^{15}~cm^{-2}}$ and $\rm{7.0 \times 10^{14}~ cm^{-2}}$, respectively. Data points below the mean threshold need to be treated with care, as they may have large error and selection effects. It should be noted that threshold describes where the distribution seen in Figure~\ref{Fig:N13N18} won't be properly sampled anymore. That could be why the curve seems to flatten off since $N({\rm C^{18}O})$ below $\rm{ 10^{15}~cm^{-2}}$ is hard to detect due to the lower line intensity. We will verify this information by increasing the sensitivity in the future.

\begin{figure}
   \centering
   \includegraphics[width=\textwidth]{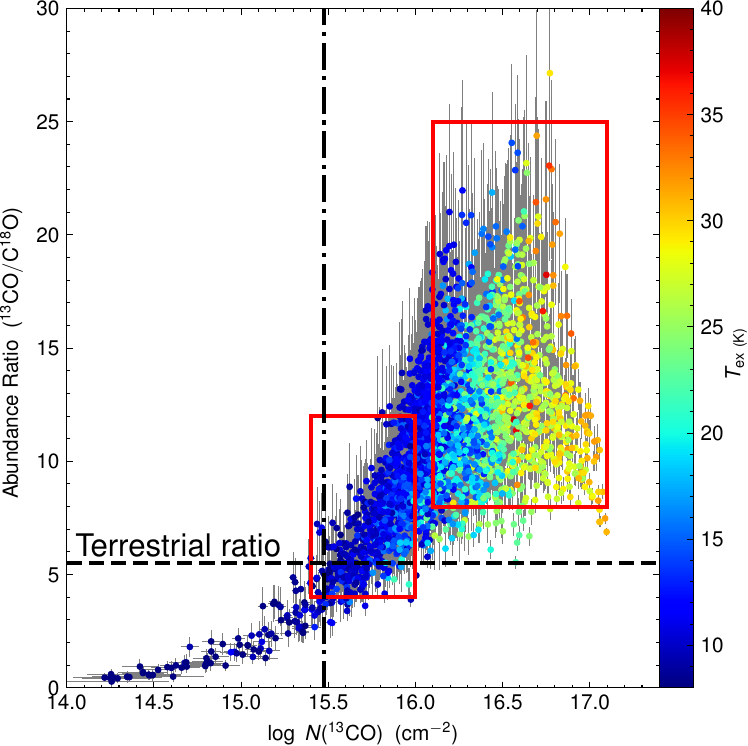}
   \caption{${N_{\rm ^{13}CO}}$ vs $X_{\rm ^{13}CO}/X_{\rm C^{18}O}$ pixel by pixel measured in 32 clouds. The color coding show the excitation temperature derived from {$^{12}$CO} peak intensity. The two dashed lines show the mean threshold of {C$^{18}$O} $\rm{7.0 \times 10^{14} cm^{-2}}$ and the terrestrial ratio $X_{\rm ^{13}CO}/X_{\rm C^{18}O}$ 5.5 (\citealt{1992A&ARv...4....1W}), respectively. The two red rectangles present two typical distribution region of type I and type II clouds classified in \citet{2019ApJS..243...25W}. The gray lines show the error bar for each pixel.}
    \label{Fig:N13abund}
\end{figure}

Figure~\ref{Fig:N13abund} illustrates the correlation between ${N_{\rm ^{13}CO}}$ and the abundance ratio ${X_{\rm ^{13}CO}/X_{\rm C^{18}O}}$ in clouds. The ratio displays a strong correlation with column density and temperature. Based on the temperature difference, there are obviously three distinct regions in the figure, corresponding to temperature intervals of [0, 14] K (cold), [14, 20] K (warm), and [20, 40] K (hot), respectively. The cold gas displays a wide range of ${N_{\rm ^{13}CO}}$ distribution, from $1~\times~10^{14}\rm{cm^{-2}}$ to $3.2 \times 10^{16}\rm{cm^{-2}}$ (minimum and maximum), while the warm and hot gases have narrower ${N_{\rm ^{13}CO}}$ ranges of [0.6, 3.2]~$\times~10^{16}\rm{cm^{-2}}$~ and~[1.6, 10]~$\times~10^{16}\rm{cm^{-2}}$, respectively. Similarly, the $X_{\rm ^{13}CO}/X_{\rm C^{18}O}$ distribution in the cold gas also has a wider range [0, 22], but with smaller dispersion, while the warm and hot gases concentrate in the range of 5 to 20 but with larger dispersion. Interestingly, the distribution of hot and cold gases is almost perfectly staggered in Figure~\ref{Fig:N13abund}, while the distribution of warm gas appears as a transition zone. In addition, as Figure~\ref{Fig:N13N18} shows, threshold describes where the distribution seen in Figure won't be properly sampled anymore. It could be why data points below the threshold have very low abundance ratios.

If we focus on the cold gas distribution in the Figure~\ref{Fig:N13abund}, the distribution follows the quadratic relation with respect to {$^{13}$CO} column density. This is probably due to isotope fractionation reactions. As demonstrated by \citet{2014MNRAS.445.4055S}, both isotopic species are effectively shielded from interstellar FUV photons in the dense region (2 mag $< \rm A_{V}< 5 mag$) due to the shielding effect of dust absorption and the increasing CO column density. This means that in the absence of additional heating sources, the ${N_{\rm ^{13}CO}}$ will increase in this region due to self-shielding against the interstellar radiation field. On the other hand, ionized carbon remains abundant in this region, making the chemical fractionation reaction (\citealt{1976ApJ...205L.165W}) important:
\begin{equation}
 \rm ^{13}C^{+} + ^{12}CO \rightleftarrows ^{12}C^{+} + ^{13}CO + \bigtriangleup E
\end{equation}

The rate coefficient for the left to right direction is $R_{\rm frac,CO,lr}$ = 2 $\times$ 10$^{-10}$ cm$^{3}$ s$^{-1}$, and for the other direction, it is $R_{\rm frac,CO,rl}$ = 2 $\times$ 10$^{-10}$ exp(-35 K/$T_{\rm gas}$) cm$^{3}$ s$^{-1}$. This suggests that the effect of chemical fractionation depends only on temperature and not on density. The gas in this region typically has a temperature of $\sim$10 K. The exothermic reaction (to the right, leading to energy release) is preferred, resulting in more production of {$^{13}$CO}. Under the dual influence of {$^{13}$CO} self-shielding and chemical fractionation, the column density of {$^{13}$CO} significantly increases in this region. However, in denser regions ($\rm A_{V} > 5 mag$), the CO chemistry is governed by non-isotope-selective reactions, and hence the column density of {$^{13}$CO} is expected to decline. Meanwhile, photons play a relatively small role and induce very few {$^{18}$O} fractionation effects as a result of CO self-shielding In the dense regions (\citealt{2005Natur.435..317L}, \citealt{2009ApJ...701..163S}, \citealt{2019MNRAS.485.5777L}).
\citet{1984ApJ...277..581L} developed the only {$^{18}$O} fractionation model which considers only a few isotopologues and only one fractionation reaction, namely
\begin{equation}
\rm  HC^{16}O^{+} + C^{18}O \rightarrow HC^{18}O^{+} + CO + 6.3~K
\end{equation}
This means that the column density of {C$^{18}$O} may be reduced due to chemical fractionation at low temperatures. In fact, we observe that the $X_{\rm ^{13}CO}/X_{\rm C^{18}O}$ ratio of cold gas increases monotonically until the extinction value is 7. This clearly shows that the effect of chemical fractionation depends only on temperature and the effect of chemical fractionation on the molecular cloud may be larger than expected.

In contrast, the distribution of hot and warm gas points in Figure~\ref{Fig:N13abund} differs from that of the cold gas. These hot and warm points exhibit relatively higher mean $X_{\rm ^{13}CO}/X_{\rm C^{18}O}$ ratios (12.5 and 11.0, respectively) than the cold gas(9.36), but as the {$^{13}$CO} column density increases, so does the temperature. Previous studies of Orion A (\citealt{2014A&A...564A..68S}) and the Gemini OB1 Molecular Cloud Complex (\citealt{2019ApJS..243...25W}) have shown that CO rare isotopologues associated with OB stars or H II regions are more effectively dissociated by FUV emission compared to CO. Based on the data presented in this paper, it is difficult to distinguish whether the observed deviation from chemical fractionation is due to increasing temperature or the effect of selective photodissociation. In future studies, we can investigate their effects on the model from different perspectives, such as changes in temperature and radiation, to better match the observed distribution of results presented in this paper.

Selective photodissociation and chemical fractionation are currently considered to be the main factors driving for CO isotopologue abundance variations. Stellar nucleosynthesis also affects the CO isotopologue abundance variations, but on larger scales, such as the Galactic chemical evolution. For example, in massive stars, extremely high temperatures due to gravitational collapse initiate the formation of {$^{12}$C} through the fusion of three alpha particles (\citealt{1995ApJS...98..617T}). on the other hand, in asymptotic giant branch (AGB) stars, {$^{13}$C} is a reaction intermediate in the carbon-nitrogen-oxygen (CNO) cycle (\citealt{1997nceg.book.....P}). The intermediate product, {$^{13}$C}, is mixed into the expanding atmosphere of the star and finally ejected into the interstellar medium after convective mixing (known as 'third dredge-up', \citealt{2004ApJ...613L..73H}). \citet{2005ApJ...634.1126M} measure the {$^{12}$C/$^{13}$C} ratios in Galactic molecular clouds by using the N = 1 - 0 transition of the CN radical. They found {$^{12}$C/$^{13}$C} has a gradient with Galactic distance which is a true indicator of Galactic chemical evolution. Due to the optical thickness of {$^{12}$CO}, we cannot directly determine its column density. It is difficult to obtain the {$^{12}$C/$^{13}$C} from the CO isotopologues alone. In this paper, we can not exclude selective nucleosynthesis which usually acts on larger scales as a potential driver for changes in the abundance of the CO isotopologues. The selective photodissociation and chemical fractionation are typically the most important to consider on cloud scale (\citealt{2014MNRAS.445.4055S}).

\citet{2019ApJS..243...25W} found that clouds can be classified into two types based on their ${X_{\rm ^{13}CO}/X_{\rm C^{18}O}}$ ratios, and pointed out that Type I clouds are likely at an earlier evolutionary stage compared to Type II clouds. Figure~\ref{Fig:N13abund} presents the typical distribution regions of these two types of clouds. The figure confirms that gases in the Type I distribution region have lower temperatures, column densities, and abundance ratios. Furthermore, gas temperature, column density, and abundance ratio gradually increase during the transition from the Type I to the Type II distribution region, suggesting a possible evolutionary sequence between the two types of gases.

As shown in Figure~\ref{Fig:N13abund}, ${X_{\rm ^{13}CO}/X_{\rm C^{18}O}}$ varies from 0 to 25 and is concentrated in the interval [5,18]. Although the abundance ratio ${X_{\rm ^{13}CO}/X_{\rm C^{18}O}}$ may locally vary by several times within the molecular cloud due to local environmental factors, the results in Figure~\ref{Fig:flux} indicate that the overall abundance ratio of the molecular cloud remains relatively constant at 9.7 and is not significantly affected by local changes in the physical environment.

\section{Discussion}
\label{Sec:6}

  \subsection{X-factor}
  \label{Sec:61}
Astronomers frequently use $^{12}$CO emission to estimate the mass of molecular gas through the X-factor (\citealt{2001ApJ...547..792D}, \citealt{2013ARA&A..51..207B}). The standard methodology assumes a simple relationship between the observed $^{12}$CO intensity $I^{^{12}\rm CO}$ and the column density of molecular gas $N(H_{2})$, as given by equation (8).

A corollary of this relation arises from integrating over the emitting area:
\begin{equation}
M_{\rm mol} = \alpha_{\rm CO} L_{\rm CO}
\end{equation}
Here, $M_{\rm mol}$ is in units of solar mass and $L_{\rm CO}$ is usually expressed in $\rm{K~km~s^{-1}~pc^{2}}$ if the distance is known. $\alpha_{\rm CO}$ is simply a mass-to-light ratio and is also referred to as the CO-to-H$_{2}$ conversion factor [e.g., for $X = 2 \times 10^{20}$ cm$^{-2}$ (K~km~s$^{-1}$)$^{-1}$, the corresponding $\alpha_{\rm CO}$ is $4.3~M_\odot$ ($\rm{K~km~s^{-1}~pc^{2}}$)$^{-1}$, \citealt{2013ARA&A..51..207B}].

On the other hand, for optically thin tracers like $^{13}$CO, there is also a relationship between the observed intensity $I^{^{13}\rm CO}$ and the column density of molecular gas $N(H_{2})$ based on the LTE assumption and certain abundance (\citealt{1991ApJ...374..540G,1997ApJ...476..781B}). Therefore, Figure~\ref{Fig:flux} suggests that the abundance $X$($^{13}$CO) may be a constant value on the scale of the molecular cloud as a whole, spanning six orders of magnitude. Additionally, Figure~\ref{Fig:flux} provides direct evidence for the existence of the CO-to-H2 conversion factor $\alpha_{CO}$ and offers a way to calculate it. However, due to the lack of distance information, we are unable to provide this value at present, and it will be investigated in our future work.

\subsection{Observational uncertainty}
  \label{Sec:62}
Sensitivity is an important parameter in observational uncertainty, which is generally determined by the RMS value of the spectrum. As we mentioned in the statistical study of observations in Sect~\ref{Sec:3}, selection effects may occur due to sensitivity thresholds, which can lead to bias in our understanding of the results. In fact, in addition to sensitivity, the beam filling factor is also an important and often overlooked factor that affects observation uncertainty. For example, beam filling factors are usually assumed to be unity (i.e., 1 in this paper) in the calculation of physical properties such as the excitation temperature, optical depth, and column density, which is inaccurate and may introduce systematic errors.

The beam filling factor $\eta$ appears as a correction factor for the brightness temperature according to equation (1). \citet{2021ApJ...910..109Y} investigated the dependence of the beam filling factor on angular resolution, sensitivity (noise levels), distances, and molecular tracers using CO images of a large-scale region ($25.8^{\circ} < l < 49.7^{\circ}$, $|b| < 5^{\circ}$) mapped by the MWISP data. They found that the beam filling factors of {$^{12}$CO} and {$^{13}$CO} are approximately unity in the Local arm ($\sim$ 1 kpc), the Sagittarius arm ($\sim$ 3 kpc), and the Scutum arm ($\sim$ 6 kpc) for giant molecular clouds, but drop to $\sim$ 0.7 and $\sim$ 0.6 in the Outer arm ($\sim$ 15 kpc), respectively. However, {C$^{18}$O} decreases significantly with distance and becomes approximately zero in the Outer arm.

\begin{figure}
   \centering

   \includegraphics[width=\textwidth, angle=0,clip=true,keepaspectratio=true,trim=0 0 0 0mm]{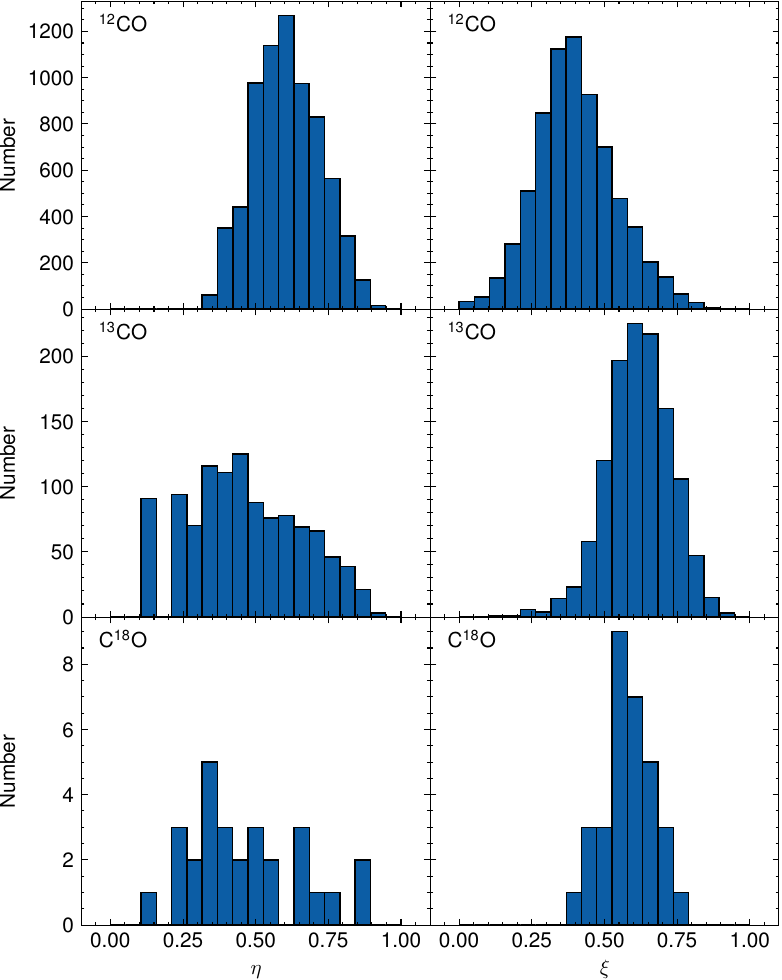}
   \caption{Histogram distribution of filling factors $\eta$ and sensitivity clip factors $\xi$ for {$^{12}$CO}, {$^{13}$CO}, and {C$^{18}$O} in the cloud catalog.}
    \label{Fig:correct}
\end{figure}

In addition to the distance factor, they found that the beam filling factor $\eta$ is mainly correlated with the angular size ${\ell}$ in beam size units and can be approximated by $0.922{\ell^2}/(\ell + 0.762)^2$ \citep{2021ApJ...910..109Y}. It should be noted that the sample of molecular clouds with {$^{13}$CO} and {C$^{18}$O} emission was relatively small when they established this relationship, which makes it likely that the error will be large for {$^{13}$CO} and {C$^{18}$O} flux and $\eta$. We need to be careful with the corrected flux of {$^{13}$CO} and {C$^{18}$O}. The angular size is defined as the equivalent diameter derived from equation (1):
\begin{equation}
\ell = \sqrt{\frac{4A}{\pi}-\Theta}
\end{equation}
where $A$ is the angular area (presented in Sect~\ref{Sec:34}) and $\Theta$ is the beam size of the MWISP survey. Figure~\ref{Fig:correct} shows the histogram distribution of beam filling factors for {$^{12}$CO}, {$^{13}$CO}, and {C$^{18}$O} in the cloud catalog. It indicates that the $\eta$ for {$^{12}$CO} is systematically greater than that for {$^{13}$CO} and {C$^{18}$O} because the emission area of {$^{12}$CO} is systematically greater than that of {$^{13}$CO} and {C$^{18}$O}.

\citet{2021ApJ...910..109Y} defined the sensitivity clip factor, $\xi$, as the ratio of the observed flux to the total flux. They also derived a relationship between the observed flux and the mean voxel S/N (definded as $x$), and the ratio of the observed flux to the total flux is approximately given by $(x - 2.224)^2/(x - 2.224 + 0.457)^2$. This relationship can be used to correct the total flux in our results.

Figure~\ref{Fig:correct} shows the histogram distribution of the sensitivity clip factor for {$^{12}$CO}, {$^{13}$CO}, and {C$^{18}$O} in the cloud catalog. It should be noted that unlike the mean voxel S/N for all the voxels in \citet{2021ApJ...910..109Y}, the mean voxel S/N of {$^{13}$CO} and {C$^{18}$O} in this paper is the mean of the S/N ratio at the peak value within all pixels. This is because the mean voxel S/N of {$^{13}$CO} and {C$^{18}$O} is so low that the calculated S/N ratio is smaller than 2.224, so we use the peak value to derive the clip factor. This results in $\xi$ values calculated for {$^{13}$CO} and {C$^{18}$O} that are generally higher than those for {$^{12}$CO}.

\begin{figure}
   \centering

   \includegraphics[width=\textwidth, angle=0,clip=true,keepaspectratio=true,trim=0 0 0 0mm]{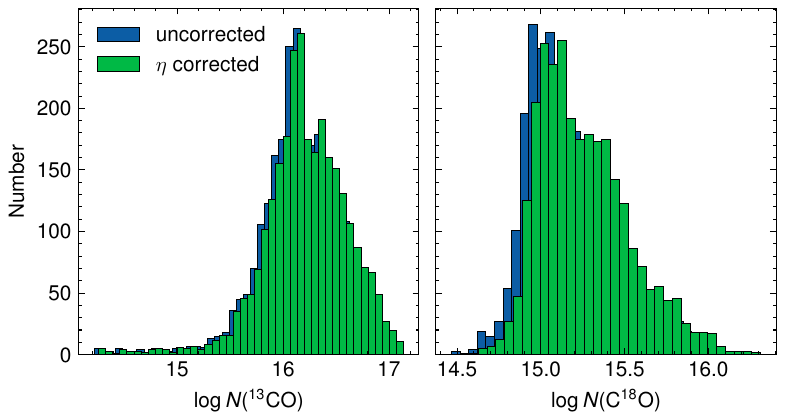}
   \caption{Histogram distribution of column density of {$^{13}$CO} and {C$^{18}$O} under LTE assumption with or without filling factor $\eta$ corrected.}
    \label{Fig:correctdensity}
\end{figure}

\begin{figure}
   \centering
   \includegraphics[width=\textwidth]{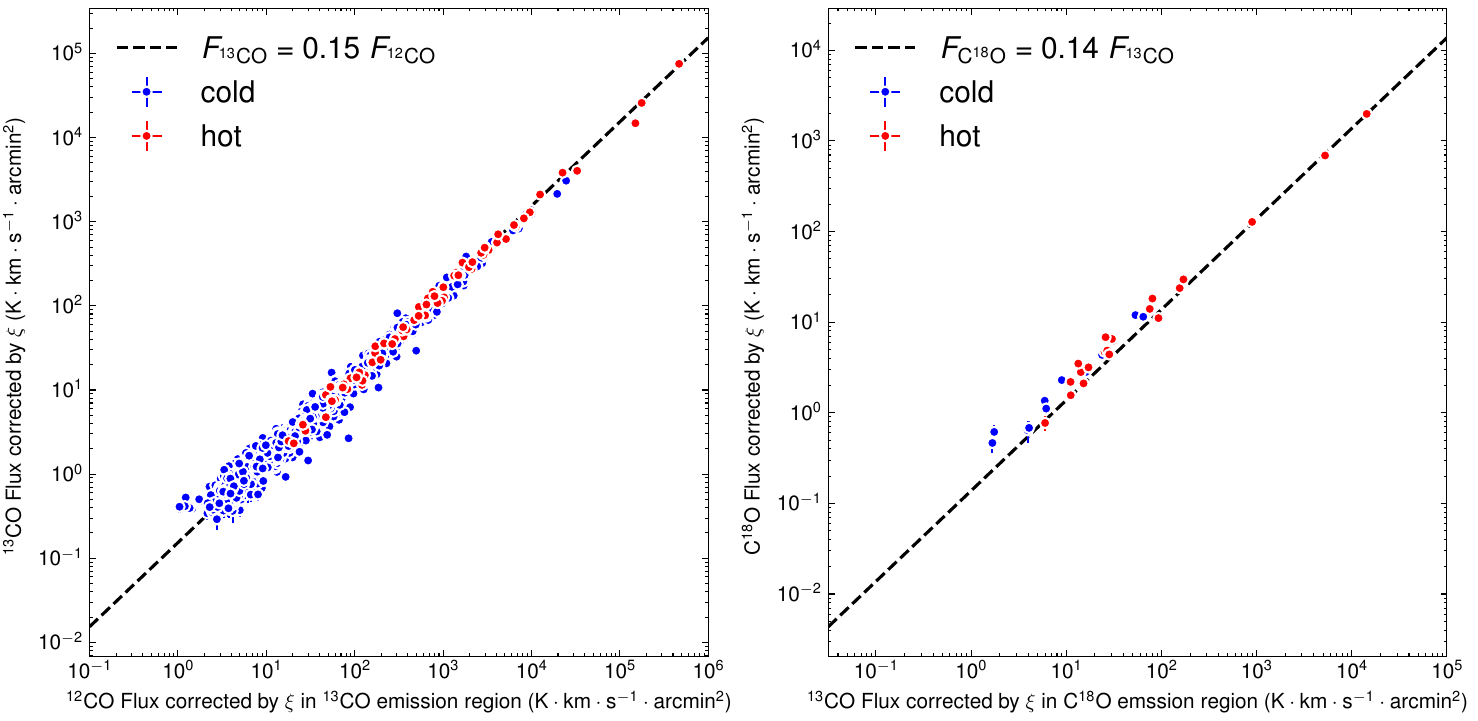}
   \caption{Left: Corrected $\rm {Flux(^{12} CO)}$ vs $\rm {Flux(^{13} CO)}$  cloud by cloud. Right: Corrected $\rm {Flux(^{13} CO)}$ vs $\rm {Flux( C^{18}O)}$ cloud by cloud. Different with Figure~\ref{Fig:flux}, the flux in here is corrected based on the sensitivity clip factor $\xi$. The colors indicate different types of molecular clouds based on excitation temperature.}
    \label{Fig:corrected-flux1}
\end{figure}

Figure~\ref{Fig:correctdensity} shows the relationship between the corrected flux based on the sensitivity clip factor $\xi$ for {$^{12}$CO} and {$^{13}$CO}, and {$^{13}$CO} and {C$^{18}$O}, respectively. Figure~\ref{Fig:corrected-flux1} also presents a significant correlation between the fluxes of CO and its isotopic molecules, and the flux ratio values are consistent with those in Figure~\ref{Fig:flux}.

The corrected relationship between CO and its isotopic molecules' fluxes are $F_{\rm ^{13}CO} = 0.15~F_{\rm ^{12}CO}$ and $F_{\rm C^{18}O} = 0.14~F_{\rm ^{13}CO}$, respectively. The derived abundance of $^{13}$CO is $5.7^{+0.1}_{-0.4} \times 10^{-7}$, and the abundance ratio $\rm X_{^{13}CO}/X_{C^{18}O}$ is $7.0^{+1.1}_{-0.1}$.

\begin{figure}
   \centering
   \includegraphics[width=\textwidth]{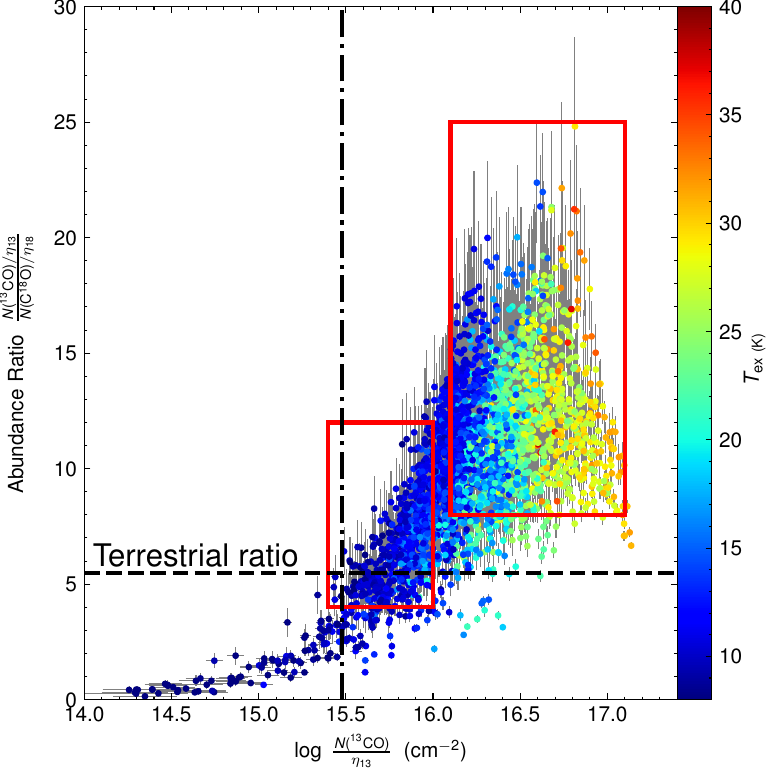}
   \caption{$\eta$ corrected ${N_{\rm ^{13}CO}}$ vs $X_{\rm ^{13}CO}/X_{\rm C^{18}O}$ pixel by pixel measured in 32 clouds. All the symbols and colors are same with Figure~\ref{Fig:N13abund}, except that the values of ${N_{\rm ^{13}CO}}$ and $X_{\rm ^{13}CO}/X_{\rm C^{18}O}$ are corrected by the fill factor $\eta$.}
    \label{Fig:N13abundcorrect}
\end{figure}

Figure~\ref{Fig:N13abundcorrect} shows histogram distribution of column density of {$^{13}$CO} and {C$^{18}$O} under LTE assumption with or without filling factor $\eta$ corrected. The $\eta$ corrected column density distribution is very close to the original column density distribution. It indicates that column density is generally underestimated without correction for this factor, but the proportion of underestimation seems to be small.

Figure~\ref{Fig:N13abundcorrect} presents the $\eta$ corrected ${N_{\rm ^{13}CO}}$ vs $X_{\rm ^{13}CO}/X_{\rm C^{18}O}$ pixel by pixel measured in 32 clouds. Its overall distribution is consistent with Figure~\ref{Fig:N13abund}, but there are some more low abundance points near the region where the {$^{13}$CO} column density is $\rm{2 \times 10^{16}~cm^{-2}}$.

\subsection{Abundance $X_{\rm C^{18}O}$}
  \label{Sec:63}

The derived abundance $X_{\rm ^{13}CO}$ and the ratio $X_{\rm ^{13}CO}/X_{\rm C^{18}O}$ of 6.5$^{+0.1}_{-0.5}$ $\times 10^{-7}$ and 9.7$^{+0.6}_{-0.8}$, respectively, suggest that the abundance of {$^{13}$CO} is higher compared to {C$^{18}$O}. Taking into account the sensitivity clip factor, the abundance of {C$^{18}$O} is corrected to be 8.1$^{+0.3}_{-0.6} \times 10^{-8}$. This value is consistent with previous studies, such as \citet{1982ApJ...262..590F} who determined $X_{\rm C^{18}O}$ for Taurus envelopes (7 $\times 10^{-8}$) and dense cores (17 $\times 10^{-8}$) based on {$^{12}$CO} (J =1-0, 2-1) and 1.3 mm continuum observations, and \citet{2002A&A...391..275H} who derived $X_{\rm C^{18}O}$ between 7 $\times 10^{-8}$ (Coalsack) and 20 $\times 10^{-8}$ (Chamaeleon I). This indicates that the derived abundance of {C$^{18}$O} in this study is within the expected range based on previous literature. As mentioned in Sect~\ref{Sec:4}, this conclusion is based on the LTE assumption and assuming that the molecular cloud has a relatively consistent optical thickness and excitation temperature. Moreover, as the correlated axes will induce a potentially bias, we remain cautious about the result that a constant abundance $X_{\rm C^{18}O}$ over the sample.

\subsection{Flux}
  \label{Sec:64}
\begin{figure}
   \centering

   \includegraphics[width=\textwidth, angle=0,clip=true,keepaspectratio=true,trim=0 0 0 0mm]{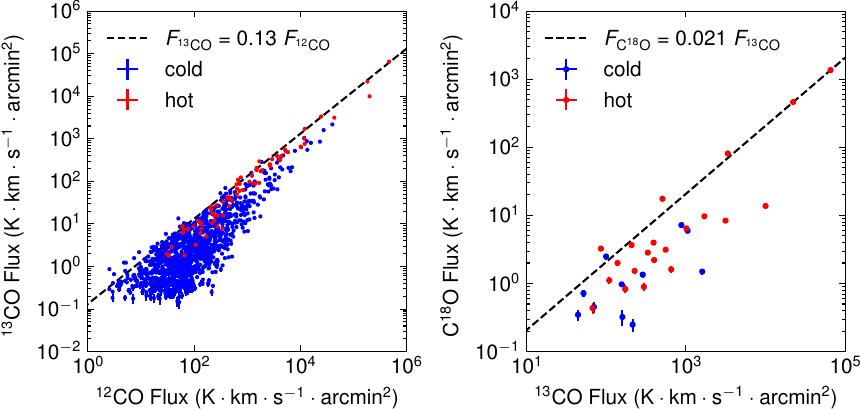}
   \caption{Left:$\rm {Flux(^{12} CO)}$ vs $\rm {Flux(^{13} CO)}$ measured cloud by cloud in 1197 clouds catalog. $\rm {Flux(^{12} CO)}$ and $\rm {Flux(^{13} CO)}$ are calculated toward the entire cloud. Right: $\rm {Flux(^{13} CO)}$ vs $\rm {Flux( C^{18}O)}$ measured in 32 clouds catalog. $\rm {Flux(^{13} CO)}$ and $\rm {Flux( C^{18}O)}$ are calculated toward the entire cloud. The colors indicate different types of molecular clouds based on excitation temperature. The dashed line shows the result of a weighted linear fit to the data.
   }
    \label{Fig:flux-all}
\end{figure}

Figure~\ref{Fig:flux} show a strong correlation between the flux of {$^{12}$CO} vs. {$^{13}$CO} and {$^{13}$CO} vs. {C$^{18}$O}, respectively. It should be noted that we keep the area the same for {$^{12}$CO} vs. {$^{13}$CO} and {$^{13}$CO} vs. {C$^{18}$O} when we calculated the flux. In fact, this linear relationships hold not only for same areas where both CO and its isotopic emissions are detected, but also for the entire molecular cloud scale. Figure~\ref{Fig:flux-all} show flux distribution of {$^{12}$CO} vs. {$^{13}$CO} and {$^{13}$CO} vs. {C$^{18}$O} calculated toward the entire cloud, respectively. Through flux-weighted linear fitting of the data, we get the relation $F_{\rm ^{13}CO} = 0.13~F_{\rm ^{12}CO}$ and $F_{\rm C^{18}O} = 0.021~F_{\rm ^{13}CO}$.
The flux ratios of {$^{13}$CO}/{$^{12}$CO} and {C$^{18}$O}/{$^{13}$CO} calculated from entire molecular clouds are slightly lower than those calculated from only the 13CO-bright and C18O-bright regions, respectively, which is foreseeable since the former calculation includes more extended areas than the latter. Although the ratios, i.e., the linear coefficients, in Figure~\ref{Fig:flux} and Figure~\ref{Fig:flux-all} are different, the {$^{12}$CO} vs. {$^{13}$CO} and {$^{13}$CO} vs. {C$^{18}$O} emissions in the two figures exhibit obvious correlations. This may imply that the flux ratio has large scale uniformity within the molecular cloud and that the flux ratio of these gases is close to the proportion within each molecular cloud. It is worth noting that the dispersion of these points in the Figure~\ref{Fig:flux-all} increases significantly with the decrease of flux. This is most likely due to sensitivity limitations. On the other hand, as shown in Sect~\ref{Sec:62}, beam filling factor is mainly correlated with the angular size. All of these can lead to underestimation of flux and cause large distribution dispersion.

\subsection{LTE assumption}
  \label{Sec:65}
As shown in the Sect~\ref{Sec:1} and Sect~\ref{Sec:3}, it is a common method to use LTE assumptions to approximate temperature, opacity, and density. However, this method relies on assumptions that might not necessarily be met in our samples. \citet{2021A&A...645A..26R} estimated the abundances, excitation temperatures, velocity field, and velocity dispersions of the three main CO isotopologues towards a subset of the Orion B molecular cloud byusing the Cramer¨CRao Bound (CRB) technique, and they found that excitation temperatures are different among {$^{12}$CO}, {$^{13}$CO}, and {C$^{18}$O}: {$^{12}$CO} presents the highest temperature, followed by {$^{13}$CO} and {C$^{18}$O}. The ratios of the estimated temperature among CO isotopologues are $T_{\rm ex}$({$^{12}$CO})/$T_{\rm ex}$({$^{13}$CO})= 1.7 and $T_{\rm ex}$({$^{13}$CO})/$T_{\rm ex}$({C$^{18}$O})= 1.3. \citet{1997astro.ph..6177P} obtain a theoretical calibration of the relation between LTE {$^{13}$CO} column density and true column density by using models of molecular clouds (MC), and non-LTE radiative transfer calculations. And, they find that LTE column density of molecular clouds typically underestimates the mean {$^{13}$CO} true column density by a factor ranging from 1.3 to 7.

In this paper, we assume that CO and its isotopologues have the same excitation temperature, which may introduce some bias to the results. So, we correct the excitation temperatures of {$^{13}$CO} and {C$^{18}$O} according to the modified formula $T_{\rm ex}$({$^{12}$CO})/$T_{\rm ex}$({$^{13}$CO})= 1.7 and $T_{\rm ex}$({$^{13}$CO})/$T_{\rm ex}$({C$^{18}$O})= 1.3 in the results of \citet{2021A&A...645A..26R}. After the correction, we find that about 40\% of the {$^{13}$CO} spectra and 3.5\% of the {C$^{18}$O} spectra have corrected excitation temperatures lower than their peak main beam brightness temperatures, which implies Equation (1) is invalid. This suggests that the modified formula for excitation temperature may not hold true for all molecular clouds. This may be because the Orion B molecular cloud \citet{2021A&A...645A..26R} studied has a more extreme environment relative to other molecular clouds, which is not common in general molecular clouds.

\begin{figure}
   \centering
   \includegraphics[width=\textwidth]{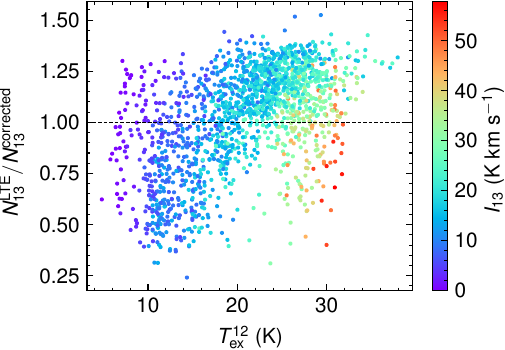}
   \caption{The ratio of ${N_{\rm ^{13}}^{LTE}}$/${N_{\rm ^{13}}^{corrected}}$ vs $T_{\rm ex}^{12}$. ${N_{\rm ^{13}}^{LTE}}$ is {$^{13}$CO} column density under LTE assumption, while ${N_{\rm ^{13}}^{corrected}}$ is corrected {$^{13}$CO} column density based on corrected {$^{13}$CO} excitation temperature and correlation $T_{\rm ex}$({$^{12}$CO})/$T_{\rm ex}$({$^{13}$CO})= 1.7. $T_{\rm ex}^{12}$ is excitation temperature from {$^{12}$CO} spectrum under LTE assumption. The color shows intensity of {$^{13}$CO}.   }
    \label{Fig:corrected-ratio13}
\end{figure}

\begin{figure}
   \centering
   \includegraphics[width=\textwidth]{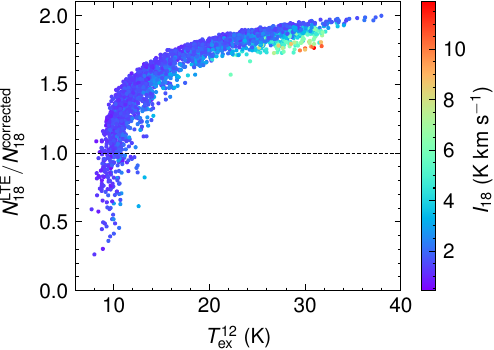}
   \caption{The ratio of ${N_{\rm ^{18}}^{LTE}}$/${N_{\rm ^{18}}^{corrected}}$ vs $T_{\rm ex}^{12}$. ${N_{\rm ^{18}}^{LTE}}$ is {$^{18}$CO} column density under LTE assumption, while ${N_{\rm ^{18}}^{corrected}}$ is corrected {$^{18}$CO} column density based on corrected {$^{18}$CO} excitation temperature and correlation $T_{\rm ex}$({$^{13}$CO})/$T_{\rm ex}$({C$^{18}$O})= 1.3. $T_{\rm ex}^{12}$ is excitation temperature from {$^{12}$CO} spectrum under LTE assumption. The color shows intensity of {$^{18}$CO}.   }
    \label{Fig:corrected-ratio18}
\end{figure}

Figure~\ref{Fig:corrected-ratio13} and Figure~\ref{Fig:corrected-ratio18} show the difference between the LTE column density and column density after excitation temperature correction from {$^{13}$CO} and {C$^{18}$O} data. For the {$^{13}$CO} LTE column density before and after the correction, their ratio ranges from 25\% to 150\% and didn't show any obvious trend with the temperature. While for the {C$^{18}$O} LTE column density before and after the correction, the ratio tend to increase with higher Tex. When $T_{\rm ex}$ $>$ 15 K, the {C$^{18}$O} LTE column density have 150\% to 200\% overestimate.

\begin{figure}
   \centering
   \includegraphics[width=\textwidth]{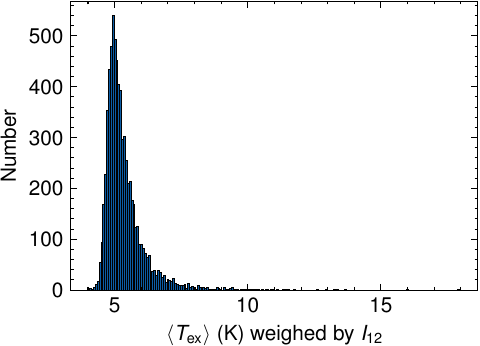}
   \caption{The histogram of the intensity-weighted average excitation temperature from {$^{12}$CO} data. }
    \label{Fig:hist-tex}
\end{figure}

\citet{2022MNRAS.510..753B} found that the intensity-weighted average excitation temperature results in the most accurate estimate of the total CO mass by simulation. Figure~\ref{Fig:hist-tex} shows the distribution of the intensity-weighted average excitation temperature in our samples.
This suggests that most molecular clouds have an overall temperature between 4 and 9 K.

\section{Summary}
\label{Sec:7}

In our study, we analyzed data from the {$^{12}$CO}, {$^{13}$CO}, and {C$^{18}$O} (1-0) transitions in a section of the third Galactic quadrant ($195^{\circ} < l < 220^{\circ}$, $|b| < 5^{\circ}$) within a velocity range of [-20, 70] km s$^{-1}$, obtained from the MWISP CO survey as described in Paper I. We compared various observational properties among the three isotopologues and summarized their parameter space based on cloud samples identified in Paper I. Our analysis revealed some new findings, and we observed that although the parameters have a wide distribution, they tend to concentrate in certain intervals according to probability density functions of the pixels:

1, The peak intensity of {$^{12}$CO}, {$^{13}$CO}, and {C$^{18}$O} is mainly concentrated in the range of [3, 8], [1.5, 4], and [0.7, 1.5] K, respectively, with mean values of 6.9, 2.0, and 1.3 K, respectively.

2, The optical depth of {$^{12}$CO}, {$^{13}$CO}, and {C$^{18}$O} is primarily distributed in the interval of [15, 40], [0.2, 0.4], and [0.05, 0.15], respectively, with mean values of 32, 0.36, and 0.12, respectively.

3, The integrated intensity of {$^{12}$CO}, {$^{13}$CO}, and {C$^{18}$O} is concentrated in the range of [5, 25], [0.4, 5], and [0.6, 2.5] K km s$^{-1}$, respectively. The majority of pixels have a ratio of ${I^{\rm ^{13}CO}/I^{^{12}\rm CO}}$ in the range of [0.1, 0.2], and a ratio of ${I^{\rm C^{18}O}/I^{\rm ^{13}CO}}$ value of approximately 0.1.

4, The angular area $A_{\rm ^{13}CO}$ of {$^{13}$CO} increases with $A_{\rm ^{12}CO}$, and its dispersion decreases. The ratio $A_{\rm ^{13}\rm CO}/A_{\rm ^{12}\rm CO}$ is 0.38 by linear fitting.

The relationship between CO and its isotope molecules flux in molecular clouds has been found to exhibit significant correlations. The first relationship is between $^{12}$CO and $^{13}$CO flux, which shows a correlation of $F_{\rm ^{13}CO}~=~0.17~F_{\rm ^{12}CO}$ across a wide range of fluxes spanning six orders of magnitude, cloud by cloud. This indicates that the abundance $X_{\rm ^{13}CO}$ is a constant for the entire samples and the X-factor, which is the ratio of CO column density to its integrated intensity, is ubiquitous in large sample molecular clouds and maintains a stable value. The derived abundance of $^{13}$CO, denoted as $X_{\rm ^{13}CO}$, is determined to be $6.5^{+0.1}_{-0.5} \times 10^{-7}$ under LTE assumption, which remains the same across the molecular cloud as a whole.

The second significant correlation is found between $^{13}$CO and C$^{18}$O flux, expressed as $F_{\rm C^{18}O}~=~0.11~F_{\rm ^{13}CO}$ cloud by cloud. This indicates that the abundance ratio $X_{\rm ^{13}CO}/X_{\rm C^{18}O}$ remains the same value of $9.7^{+0.6}_{-0.8}$ across the molecular cloud as a whole under LTE assumption. The derived abundance of C$^{18}$O, denoted as $X_{\rm C^{18}O}$, is determined to be $6.7^{+0.7}_{-0.9} \times 10^{-8}$.

The linear relationships of $F_{\rm ^{12}CO}$ vs. $F_{\rm ^{13}CO}$ and $F_{\rm ^{13}CO}$ vs. $F_{\rm C^{18}O}$ holds not only for the {$^{13}$CO}-bright region or {C$^{18}$O}-bright region, but also for the entire molecular cloud scale with lower flux ratio.

After considering observation uncertainties, the corrected relationships between CO and its isotope molecules flux are found to be $ F_{\rm ^{13}CO}~=~0.15~F_{\rm ^{12}CO}$ and $F_{\rm C^{18}O}~=~0.14~F_{\rm ^{13}CO}$, respectively. This leads to a derived abundance of $X_{\rm ^{13}CO}$ as $5.7^{+0.1}_{-0.4} \times 10^{-7}$ and an abundance ratio $X{\rm ^{13}CO}/X_{\rm C^{18}O}$ as $7.0^{+1.1}_{-0.1}$. The abundance of $X_{\rm C^{18}O}$ is determined to be $8.1^{+0.3}_{-0.6} \times 10^{-8}$.

The abundance ratio $X_{\rm ^{13}CO}/X_{\rm C^{18}O}$ shows a strong correlation with column density and temperature. The value of $X_{\rm ^{13}CO}/X_{\rm C^{18}O}$ is generally higher than the terrestrial ratio of 5.5, indicating that the chemical fractionation, selective dissociation, and self-shielding effects play an important role in determining the isotopic abundance ratios in molecular clouds. Although local environmental differences may cause some variation in $X_{\rm ^{13}CO}$ and $X_{\rm ^{13}CO}/X_{\rm C^{18}O}$ within the molecular cloud, on a global scale, they both maintain a stable value cloud by cloud and are independent of evolution. This implies that while local properties of molecular clouds may be affected by star-forming activities during the evolutionary process, such as changes in temperature, density, and abundance, the overall properties of molecular clouds do not change significantly.

\begin{acknowledgements}
This research made use of the data from the Milky Way Imaging Scroll Painting (MWISP) project, which is a multi-line survey in {$^{12}$CO}/{$^{13}$CO}/{C$^{18}$O} along the northern galactic plane with PMO-13.7m telescope. We are grateful to all the members of the MWISP working group, particularly the staff members at PMO-13.7m telescope, for their long-term support. MWISP was sponsored by National Key R$\&$D Program of China with grant 2017YFA0402701 and CAS Key Research Program of Frontier Sciences with grant QYZDJ-SSW-SLH047. This work is supported by the Natural Science Foundation of Jiangsu Province(Grants No BK20201108). This work is supported by the National Natural Science Foundation of China (grants No. 12073079). J.Y. is supported by National Natural Science Foundation of China through grant 12041305. C.W. thanks to the supported by China Postdoctoral Science Foundation (No. 2022M713173). Y.Su is supported by the National Natural Science Foundation of China (grants No. 12173090). We also thanks to the support by Millimeter Wave Radio Astronomy Database (http://www.radioast.csdb.cn/). C.W. is grateful for the funding of the FAST Fellow. We would like to thank the referee for going through the paper carefully and much appreciate the many constructive comments that improved this paper.

\end{acknowledgements}

%






\bibliography{bibtex}{}
\bibliographystyle{aasjournal}



\clearpage

\begin{longrotatetable}
\begin{deluxetable*}{cDDccDDDDDDDDDc}
\tablecaption{A catalog of Observational properties \label{table1}}
\tablewidth{700pt}
\tabletypesize{\scriptsize}
\tablehead{
  \colhead{} &
  \multicolumn2c{$T_{\mathrm{peak}}^{12}$} &
  \multicolumn2c{$T_{\mathrm{peak}}^{13}$} &
  \colhead{$T_{\mathrm{peak}}^{18}$} &
  \colhead{$T_{\mathrm{ex}}$}  &
  \multicolumn2c{$\tau_{12}$} &
  \multicolumn2c{$\tau_{13}$} &
  \multicolumn2c{$\tau_{18}$} &
  \multicolumn2c{$I_{12}$} &
  \multicolumn2c{$I_{13}$} &
  \multicolumn2c{$I_{18}$} &
  \multicolumn2c{$A_{12}$} &
  \multicolumn2c{$A_{13}$} &
  \multicolumn2c{$A_{18}$} & \colhead{} \\
  \colhead{}    &
  \multicolumn2c{(K)} &
  \multicolumn2c{(K)} &
  \colhead{(K)} &
  \colhead{(K)} &
  \multicolumn2c{}    &
  \multicolumn2c{}    &
  \multicolumn2c{}    &
  \multicolumn2c{$(\mathrm{K~km~s^{-1}})$} &
  \multicolumn2c{$(\mathrm{K~km~s^{-1}})$} &
  \multicolumn2c{$(\mathrm{K~km~s^{-1}})$} &
  \multicolumn2c{$(\mathrm{arcmin^2})$} &
  \multicolumn2c{$(\mathrm{arcmin^2})$} &
  \multicolumn2c{$(\mathrm{arcmin^2})$}  & \colhead{} \\
}
\decimals
\startdata
{\bf Min} & 1.4759 & 0.6658 & 0.7759 & 4.5090 & 4.8559 & 0.0546 & 0.0279 & 0.8614 & 0.3821 & 0.5701 & 1.0000 & 0.5000 & 0.5000 & \\
{\bf Max} & 37.5405 & 15.1960 & 3.5308 & 41.0634 & 309.9523 & 3.4826 & 0.6116 & 206.2471 & 57.9293 & 11.6475 & 35751.7500 & 18171.0000 & 1028.0000 & \\
{\bf Mean} & 6.8713 & 1.9809 & 1.3343 & 10.1883 & 32.0960 & 0.3606 & 0.1190 & 19.1297 & 3.3174 & 1.9735 & 44.6266 & 45.3179 & 48.9141 &\\
\enddata
\tablecomments{Columns are peak intensity, excitation temperature, optical depth, integrated intensity, and emission are for {$^{12}$CO}, {$^{13}$CO}, and {C$^{18}$O}, respectively. The rows are the minimum, maximum, and mean values of the corresponding parameters}
\end{deluxetable*}
\end{longrotatetable}

\clearpage

\clearpage

\clearpage

\clearpage

\clearpage

\clearpage

\clearpage

\clearpage

\clearpage

\clearpage

\clearpage

\clearpage

\begin{figure}
   \centering
   \includegraphics[width=\textwidth]{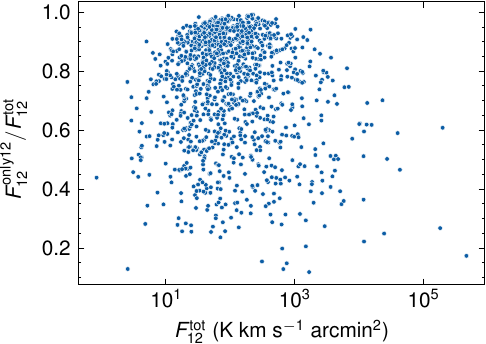}
   \caption{{$^{12}$CO} flux $\rm {F_{12}^{tot}}$ vs flux proportion of {$^{12}$CO}-only region in 1197 clouds catalog. $\rm {F_{12}^{tot}}$ is calculated toward the entire cloud.}
    \label{Fig:corrected-flux}
\end{figure}

\clearpage

\clearpage

\clearpage

\clearpage

\end{document}